\documentclass[12pt]{article}
\usepackage{epsfig,amsfonts,amsthm}
\newcommand{\be}{\begin{equation}}
\newcommand{\ee}{\end{equation}}
\newcommand{\bea}{\begin{eqnarray}}
\newcommand{\eea}{\end{eqnarray}}
\newcommand{\dst}{\displaystyle}
\newcommand{\fr}[2]{\frac{{\dst #1}}{{\dst #2}}}
\newcommand{\f}{\phi}
\newcommand{\fd}{\phi^\dagger}
\renewcommand{\Re}{\mbox{Re }}
\renewcommand{\Im}{\mbox{Im }}
\newcommand{\stolb}[3]{ \left( \begin{array}{c}#1 \\ #2 \\ #3\end{array}\right) }
\newcommand{\stolbik}[2]{ \left( \begin{array}{c}#1 \\ #2 \end{array}\right) }
\newcommand{\bracket}[2]{ \langle #1|#2 \rangle}

\newtheorem{theorem}{Theorem}
\newtheorem{proposition}[theorem]{Proposition}

\def\lsim{\mathrel{\rlap{\lower4pt\hbox{\hskip1pt$\sim$}}
    \raise1pt\hbox{$<$}}}         
\def\gsim{\mathrel{\rlap{\lower4pt\hbox{\hskip1pt$\sim$}}
    \raise1pt\hbox{$>$}}}         

\title{Minkowski space structure of the Higgs potential in 2HDM}

\author{I.P. Ivanov\thanks{E-mail: Igor.Ivanov@ulg.ac.be}\\
  {\small Istituto Nazionale di Fisica Nucleare, Gruppo collegato di Cosenza,} \\
  {\small I-87036 Arcavacata di Rende, Cosenza, Italy,}\\
  {\small Physique th\'{e}orique fondamentale, D\'{e}p. de Physique, Universit\'{e} de Li\`{e}ge,} \\
  {\small All\'{e}e du 6 Ao\^{u}t 17, b\^{a}t. B5a, B-4000 Li\`{e}ge, Belgium}\\
  {\small and}\\
  {\small Sobolev Institute of Mathematics, ac. Koptyug av. 4, 630090, Novosibirsk, Russia}}

\begin{document}
\maketitle

\begin{abstract}
The Higgs potential of 2HDM keeps its generic form under the group
of transformation $GL(2,C)$, which is larger than the usually considered
reparametrization group $SU(2)$. This reparametrization symmetry
induces the Minkowski space structure in the orbit space of 2HDM.
Exploiting this property, we present a geometric analysis of the number and properties
of stationary points of the most general 2HDM potential.
In particular, we prove that charge-breaking and neutral vacua never coexist in 2HDM
and establish conditions when the most general explicitly $CP$-conserving
Higgs potential has spontaneously $CP$-violating minima.
We also define the prototypical model of a given 2HDM, which has six free parameters less
than the original one but still contains all the essential physics.
Our analysis avoids manipulation with high-order algebraic equations.
\end{abstract}

\section{Introduction}

The Standard Model relies on the Higgs mechanism of the
electroweak symmetry breaking. Its simplest
realization is based on a single weak isodoublet of scalar fields,
which couple to the gauge and matter fields and self-interact
via the quartic potential, for review see \cite{Hunter,djouadi1}.
Extended versions of the Higgs mechanisms are based on
more elaborate scalar sectors.
The two-Higgs-doublet model \cite{2HDM},
where one introduces two Higgs doublets $\phi_1$
and $\phi_2$, is one of the most economic
extensions of the Higgs sector beyond the Standard Model.
This model has been extensively studied in literature from
various points of view, see \cite{Hunter,Sanchez,ginzreview,haber} and references
therein. The minimal supersymmetric extension of the Standard Model (MSSM)
uses precisely a specific version of the 2HDM to break the electroweak symmetry,
\cite{djouadi2}.

The Higgs potential of the most general 2HDM $V_H = V_2 + V_4$ is
conventionally parametrized as
\bea
V_2&=&-{1\over 2}\left[m_{11}^2(\phi_1^\dagger\phi_1) + m_{22}^2(\phi_2^\dagger\phi_2)
+ m_{12}^2 (\phi_1^\dagger\phi_2) + m_{12}^{2\ *} (\phi_2^\dagger\phi_1)\right]\,;\nonumber\\
V_4&=&\fr{\lambda_1}{2}(\phi_1^\dagger\phi_1)^2
+\fr{\lambda_2}{2}(\phi_2^\dagger\phi_2)^2
+\lambda_3(\phi_1^\dagger\phi_1) (\phi_2^\dagger\phi_2)
+\lambda_4(\phi_1^\dagger\phi_2) (\phi_2^\dagger\phi_1) \label{potential}\\
&+&\fr{1}{2}\left[\lambda_5(\phi_1^\dagger\phi_2)^2+
\lambda_5^*(\phi_2^\dagger\phi_1)^2\right]
+\left\{\left[\lambda_6(\phi_1^\dagger\phi_1)+\lambda_7
(\phi_2^\dagger\phi_2)\right](\phi_1^\dagger\phi_2) +{\rm
h.c.}\right\}\,,\nonumber
\eea
with 14 free parameters: real $m_{11}^2, m_{22}^2, \lambda_1, \lambda_2, \lambda_3, \lambda_4$
and complex $m_{12}^2, \lambda_5, \lambda_6, \lambda_7$.
The large number of free parameters makes the analysis of the most general 2HDM
and its phenomenological consequences rather complicated. It suffices to say
that even the equations for the stationary points and the mass matrices of the most general 2HDM are too
cumbersome to solve them explicitly or to extract any useful information in a straightforward way.
Appendix A of \cite{Sanchez} serves as a remarkable illustration of this point.

Fortunately, in many phenomenological applications one does not actually need to consider
the {\em most} general 2HDM. For example, the spontaneous $CP$-violation takes place in the Higgs sector
of 2HDM even with $m_{12}^2 = 0$ and real $\lambda_5$, $\lambda_6$, $\lambda_7$, as originally
studied in \cite{2HDM}. In an even simpler case $m_{12}^2 = \lambda_5 = \lambda_6 = \lambda_7 = 0$
there is still much room for very rich phenomenology since the two doublets can couple to fermions
in different ways. The MSSM is also based
upon a very specific version of 2HDM with $\lambda_1 = \lambda_2 = -2\lambda_3$
and $\lambda_5 = \lambda_6 = \lambda_7 = 0$.
In all these cases the straightforward algebra is usually sufficient for the complete treatment
of the EWSB.

Still, there are issues that go beyond these simple versions of 2HDM and call upon
the study of the most general 2HDM.
First, one might be interested in understanding {\em what can} and {\em what cannot} happen,
in principle, if the Higgs mechanism involves two doublets. Next, when discussing
{\em natural} sets of the free parameters of the Higgs potential, one runs
into the question of {\em crucial} and, in some sense, {\em redundant} parameters.
In other words, one wants to know the ``space of different physical situations"
realized in 2HDM, not the space of different free parameters.
Attempting to link different-looking Higgs potentials with
the same physical content with some reparametrization transformation (see below),
one unavoidably plunges into the analysis of the most general 2HDM.
Finally, one can go even further and consider more evolved Higgs sectors, like 3HDM,
where the straightforward algebra strongly limits one's insight into phenomenological
possibilities.

The problem that emerges is not to calculate precisely
the vacuum expectation values (v.e.v.'s) of the fields, the Higgs masses etc.,
but to {\em understand the general structure in the space of all 2HDM's}.

\subsection{Reparametrization invariance}

For a long time, the knowledge of this structure had been very limited. One would set some parameters
to zero, and occasionally observe that some of the remaining ones were ``more important" than the others.
The situation began to change several years ago with the observation of the
{\em reparametrization freedom} and the development of the {\em basis-independent methods},
\cite{haber,CP}.

In general, a theory has reparametrization invariance if two different sets of parameters
lead to the same observables. In 2HDM one has the Higgs potential (\ref{potential}) that depends on Higgs fields $\phi_\alpha$
and coupling constants $\lambda_i,\,m_{ij}^2$, $V = V(\lambda_i,\,m_{ij}^2;\,\phi_\alpha)$. The physical observables
are functions of these coupling constants, but not fields themselves.
The key observation is that if one performs a reparametrization of the potential,
\be
\lambda_i,\,m_{ij}^2 \to \tilde{\lambda}_i,\,\tilde{m}_{ij}^2\,,\label{rep}
\ee
then the resulting Higgs potential $V(\tilde{\lambda}_i,\,\tilde{m}_{ij}^2;\,\phi_\alpha)$,
via a transformation of the Higgs fields
$$
\phi_\alpha \to \tilde{\phi}_\alpha = R_{\alpha\beta}\phi_\beta\,,
$$
can in some cases be brought back to the form with original coefficients
$V(\lambda_i,\,m_{ij}^2;\,\tilde{\phi}_\alpha)$. When this happens, the essential physics
encoded in $V(\lambda_i,\,m_{ij}^2;\,\phi_\alpha)$ and $V(\tilde{\lambda}_i,\,\tilde{m}_{ij}^2;\,\phi_\alpha)$
is the same. In other words, if one performs transformation (\ref{rep}) and, simultaneously, inverse
transformation of the fields, $\phi_\alpha \to (R^{-1})_{\alpha\beta}\phi_\beta$,
one obtains exactly the same expression for the potential as before.

By construction, the Higgs fields and the coupling constants live in different spaces.
It is convenient to think of the Higgs potential as a scalar product between the ``tensors"
composed of fields and ``tensors" composed of coupling constants. The initial form of potential
corresponds to the explicit coordinate representation of these tensors in some specific basis.
Rotating the basis, one induces the above transformations of the fields and of the coordinates,
which leaves the potential and the physics invariant.

The reparametrization symmetry offers the freedom to choose the basis for description
of the Higgs potential. The quantities that are invariant under basis rotations reflect the physics,
while quantities that depend on the basis transformations contain redundant information
about the approach chosen.

The standard way to implement the reparametrization freedom is to consider the doublets $\phi_1$ and $\phi_2$
as up and down components of a ``hyperspinor" $\Phi$ and
perform a unitary transformation $\Phi \to U\Phi$, $U \in U(2)$.
All the 2HDM's whose potentials are related by unitary transformations from $U(2)$
correspond to the same physical theory written in different bases.
Obviously, for the basis-independent analysis, one must use the most general form of the potential.

Representing the Higgs potential as
\be
V = Y_{ab} (\Phi^\dagger_a \Phi_b) + {1\over 2}Z_{abcd}(\Phi^\dagger_a \Phi_b)(\Phi^\dagger_c \Phi_d)\,,
\label{potentialYZ}
\ee
with tensors $Y_{ab}$ and $Z_{abcd}$ being the short notation for the mass terms and quartic couplings,
respectively, one can construct from them many new tensors. The complete traces of these tensors
are the $SU(2)$ invariants. The physical observables must be constructed from these invariants.
The parameters that do change under the $SU(2)$
transformation are redundant; they depend on the way we describe the model
but not on the model itself. A heavy machinery can be developed along these lines,
see \cite{haber} for details.

\subsection{Discovering structures in the space of two-Higgs-doublet models}

Representation (\ref{potentialYZ}), although compact, sheds little light on the structure
of the space of all 2HDM and, therefore, on the properties of a generic model.
For example, a great calculational effort was undertaken in \cite{haber2} to establish
the complete set of basis-independent conditions for the explicit $CP$-conservation
in the most general 2HDM. This analysis revealed four real invariants,
but the meaning and origin of them remained a mystery.

The structure of the space of all 2HDM became clearer after \cite{ivanov},
where a simple yet efficient group-theoretic analysis of the 2HDM was performed.
Tensors $Y_{ab}$ and $Z_{abcd}$ were decomposed into irreducible representations
of $SU(2)$ and then were mapped onto the irreducible representations of the group
$SO(3)$. The complicated tensorial analysis was
reduced to the simple linear algebra in 3D linear vector space
with one real symmetric matrix and two real vectors.
Basis invariant conditions for the hidden $Z_2$ symmetry and for the explicit
$CP$-conservation could be easily derived in these linear-algebraic terms.

Recently, a similar approach was used in \cite{nishi} to extend
such geometric analysis to the models with $N$ Higgs doublets.
From the very beginning, the author of \cite{nishi} switches from the
fundamental to the adjoint representation of $SU(2)$, or $SU(N)$,
and derives the criteria for the absence of explicit
and spontaneous $CP$-violation in these models.
Another realization of such analysis was presented of \cite{maniatis}.
Authors of this work went on to find and study the stationary points of the 2HDM potential
using interesting algebraic techniques.
The authors of another recent paper
\cite{barroso2006} also studied a generic NHDM,
aiming at discovering general properties and the underlying structure of this model.

It is interesting to note that back in 1970-1980's there was much activity on
mathematical properties of various realizations of the
symmetry breaking Higgs mechanisms. It was understood that
the problem of minimization of some group-invariant potential is simplified
if one switches from the space of Higgs fields to the {\em orbit space}, \cite{michel}.
This idea was exploited in \cite{Kim} and in \cite{sartori} to study the minima of
a Higgs potential invariant under the Lie group $G$ with Higgs fields transforming
under various representations of this group. In electroweak theory,
$G = SU(2)\times U(1)_{EW}$, but this simple case has never been analyzed
in detail under the most general circumstances. For a recent application of these old
ideas to 2HDM see \cite{sartori2hdm}. \\

The recent analyses as well as the older mathematical work
make it clear that in spite of the messy parametrization (\ref{potential}),
the Higgs potential in 2HDM must have a very simple internal structure. In this paper
we make a further step to uncover and put to work this structure.

In Section 2 we note that the Higgs potential (\ref{potential}) keeps its generic form
under the action of a larger group of transformations between the two doublets, $GL(2,C)$.
This leads us to the group $SO(1,3)$ of the transformation of the adjoint representation, and
induces the Minkowski space structure in the orbit space of 2HDM.
Even more important, it induces the Minkowski space structure just {\em in the space of
2HDM's} themselves. We conclude Section 2 by defining a {\em prototypical model}
of a given generic 2HDM, which involves six parameters less but still contains all the essential physics.

We then use this formalism to study the existence and the number
of the stationary points that correspond
to the charge-breaking (Section 3) and neutral (Section 4) vacua,
as well as the mass matrices at these points.
We do this in geometric terms, avoiding explicit manipulation
with high-order algebraic equations. Then, in Section 5
we derive a criterion when the most general 2HDM with explicit $CP$-conservation
has spontaneously $CP$-violating solutions.
Finally, in Section 6 we draw conclusions. Three Appendices provide additional details.

\section{Minkowski space of 2HDM}
\subsection{Extending the reparametrization group}

The Higgs potential (\ref{potential}), defined in the 8-dimensional space of Higgs fields,
depends on them through their
electroweak-invariant combinations (EW-scalars) $x_{ij}=(\fd_i \f_j)$, $i,j = 1,2$.
The points in the space of Higgs fields that have the same values
of $x_{ij}$ can be related by a unitary transformation from the electroweak group
$SU(2)\times U(1)_{EW}$ and therefore form an orbit.
Points within each orbit are indistinguishable for the Higgs potential, and
one can think of the Higgs potential as defined in the {\em orbit space}.

Since the two Higgs doublets $\phi_1$ and $\phi_2$ in 2HDM have the same quantum numbers,
we treat them as the up and down components of a single
contravariant hyperspinor $\Phi = (\phi_1,\, \phi_2)^T$, which realizes
the fundamental representation of $SU(2)$. The quartic part of the potential
can be viewed as a twice-covariant, twice-contravariant tensor, with
coefficients $\lambda_i$ being its coordinates in a chosen basis.
The rotation from one basis to another is realized by a unitary transformation from
$SU(2)$, which induces the corresponding transformation
both of the Higgs fields $\Phi \to \Phi'$
(and, consequently, the basis tensors) and of the coordinates, $\lambda \to \lambda'$,
leaving the physics invariant, see details in \cite{ivanov}.
This reparametrization group offers freedom in describing the Higgs potential and,
in general, the entire 2HDM.

It is very convenient to switch from the fundamental to adjoint representation of $SU(2)$,
\cite{nishi}.
As $2\otimes \bar 2 = 3\oplus 1$, the quantities $x_{ij}$ form a singlet and a triplet. Using the well-known
$SU(2)\to SO(3)$ mapping, one maps them into a scalar and a real-valued vector:
\be
r_0 = (\Phi^\dagger \Phi) = (\fd_1 \f_1) + (\fd_2 \f_2)\,,\quad
r_i = (\Phi^\dagger \sigma_i \Phi) =
\stolb{(\fd_2 \f_1) + (\fd_1 \f_2)}{-i[(\fd_1 \f_2) - (\fd_2 \f_1)]}{(\fd_1 \f_1) - (\fd_2 \f_2)}\,.
\label{ri}
\ee
The Higgs potential can be now written as
\be
V = - M_i r_i - M_0 r_0 + A_{ij} r_i r_j + B_i r_i r_0 + C r_0^2\,,\label{potentialABC}
\ee
where $M_i$ and $M_0$ contain the mass coefficients, while $A_{ij}$, $B_i$ and $C$ are composed
of quartic couplings $\lambda_i$, see details in \cite{ivanov,nishi,maniatis}.

Representation (\ref{potentialABC}) already displays some structure in the space of all possible
2HDM, i.e. the space of all free parameters of the potential.
An $SU(2)$ rotation induces a corresponding $SO(3)$ rotation
of the basis in 3D space, under which $r_i$, $M_i$, $B_i$ transform as vectors, $A_{ij}$
transforms as a symmetric tensor, but the value of the potential remains the same.
The physically relevant quantities, the scalars of $SO(3)$, are not just these vectors and tensors but
their scalar products and traces.\\

We now make a step forward and consider the largest group of invertible linear transformations $GL(2,C)$ of a complex-valued 2-vector.
Since, for instance, the quartic part of the Higgs potential contains all possible fourth order
terms, an arbitrary linear transformation between the two doublets
keeps it intact, only up to reparametrization. Therefore, the group under which the Higgs potential
is reparametrization-invariant is $GL(2,C)$, not just $U(2)$ as usually assumed.

The subgroup $C^*$ (i.e. overall multiplication by a non-zero complex number) of $GL(2,C)$, generated by a unit operator,
is redundant for the description of the potential. Multiplication of all the fields by the same real non-zero constant leads to
rescaling of all the observables, without changing the structure of the model,
while the global phase rotations have no effect on the potential.
It is its factorgroup $SL(2,C)$ that embraces all non-trivial transformations
and generates interesting symmetries.

For the adjoint representation, $SL(2,C)$ induces the proper Lorentz group $SO(1,3)$.
Apart from the 3D rotations, induced by $SU(2)$, one also has boosts along the three axes.
The scalar $r_0$ and vector $r_i$ now become parts of a single irreducible representation of
$SO(1,3)$:
\be
r^{\mu} = (r_0, r_i)\,.
\ee
Thus, the orbit space, which is given by all possible four-vectors $r^\mu$, is equipped with
the Minkowski space structure. The covariant and contravariant vectors are related, as usual, by $g_{\mu\nu} = \mathrm{diag}(1,-1,-1,-1)$.
For convenience, when describing geometry of this space,
we will coin the standard space-time terms ``future", ``past", ``time-like" and ``space-like", whose definitions are obvious.

Now, the Higgs potential in the orbit space can be written in a very compact form:
\be
V = - M_\mu r^\mu + {1\over 2}\Lambda_{\mu\nu} r^\mu r^\nu\,,\label{Vmunu}
\ee
where
\bea
M^\mu &=& {1\over 4}\left(m_{11}^2+m_{22}^2,\, -2\Re m_{12}^2,\,
2\Im m_{12}^2,\, -m_{11}^2+m_{22}^2\right)\,,\nonumber\\[2mm]
\Lambda^{\mu\nu} &=& {1\over 2}\left(\begin{array}{cccc}
{\lambda_1+\lambda_2 \over 2} + \lambda_3 & -\Re(\lambda_6 + \lambda_7)
    & \Im(\lambda_6 + \lambda_7) & -{\lambda_1-\lambda_2 \over 2} \\[1mm]
-\Re(\lambda_6 + \lambda_7) & \lambda_4 + \Re\lambda_5 & -\Im\lambda_5 & \Re(\lambda_6 - \lambda_7) \\[1mm]
\Im(\lambda_6 + \lambda_7) & -\Im\lambda_5 & \lambda_4-\Re\lambda_5 & -\Im(\lambda_6 - \lambda_7) \\[1mm]
 -{\lambda_1-\lambda_2 \over 2} & \Re(\lambda_6 - \lambda_7) & -\Im(\lambda_6 - \lambda_7)
 & {\lambda_1+\lambda_2 \over 2} - \lambda_3
\end{array}\right)\,.\label{lambda}
\eea
This way of presenting the potential appeared in \cite{nishi}; however index $\mu$
was assumed to be just a short notation for a pair ``singlet + triplet". The wider group
of spinor transformation was not exploited.

\subsection{Extended reparametrization group and kinetic term}\label{sect-kinetic}

The extended reparametrization group $SL(2,C)$ modifies the Higgs kinetic term. Introducing
\be
\rho^\mu = (\partial_\alpha \Phi)^\dagger \sigma^\mu (\partial^\alpha \Phi)\,,
\ee
where $\alpha$ indicates the true space-time coordinates, one can rewrite the kinetic term as
\be
K = K_{\mu}\rho^\mu\,.
\ee
In the standard form, $K_{\mu} = (1,\, 0,\, 0,\, 0)$,
which sets a preferred ``reference frame" in the space of $\rho^\mu$ and, therefore,
in the orbit space.
Upon an $SL(2,C)$ transformation of the Higgs fields, spacelike components of $K_{\mu}$
become non-zero, and the canonical form of the kinetic part is broken.

The structure of the kinetic term does not affect the number of the stationary points
of the potential, however it does affect the position of these stationary points and
the matrices of second derivatives calculated around them.
If $\varphi_a$ are the neutral scalar fields, then the second derivative matrix
\be
M_{ab} = {1\over 2} {\partial^2 V \over \partial \varphi_a\,\partial \varphi_b}\,,\quad a=1,\,\dots,\,8\,,
\ee
represents the true mass matrix of the Higgs bosons
only when calculated in the above preferred ``reference frame".
However, the true mass matrix and the second derivative matrix in any Lorentz frame
share a common property: the positive definiteness.
\begin{proposition}\label{prop-positive-definite}
Any $SL(2,C)$ transformation of the Higgs fields preserves the positive (semi)-definiteness
of the second derivative matrix.
\end{proposition}
\begin{proof}
Consider each of the complex Higgs fields as a pair of real Higgs fields. An $SL(2,C)$ transformation
induces an $SL(4,R)$ transformation of the resulting eight real Higgs fields $\varphi_a$:
$$
\varphi_a \to \varphi_a^\prime = R_{aa'}\varphi_{a'}\,.
$$
The mass matrix is then transformed as
$$
M_{ab} \to M_{ab}^\prime = M_{a'b'}(R^{-1})_{a'a}(R^{-1})_{b'b}\,.
$$
Suppose $M_{ab}^\prime$ is positive (semi)definite; then $M_{ab}^\prime q_a q_b > 0$
(or $\ge 0$) for any vector $q_a$.
Denote $Q_a = q_{a'} (R^{-1})_{a'a}$. Since map $R_{a'a}$ is surjective,
then $Q_a Q_b M_{ab} > 0$ (or $\ge 0$) for any $Q_a$. Therefore, $M_{ab}$ is also positive (semi)definite.
\end{proof}
This proposition states nothing but the fact that a local minimum remains a local minimum under
any rotation and stretching of the coordinates in the space of Higgs fields.

Thus, the fact that kinetic term is not conserved under $SL(2,C)$ transformations
has little relevance to the study of the overall properties of 2HDM.
The formalism to be developed here is particularly useful to study the following issues:
\begin{enumerate}
\item the number of stationary points of the potential and their charge-breaking/neutral nature;
\item the minimum/saddle point classification of the stationary points and the number of the Goldstone modes;
\item spontaneous $CP$-violation in an explicitly $CP$-conserving 2HDM;
\item possible phase transitions under continuous variation of the parameters of the potential.
\end{enumerate}
In the present paper we pay attention to the first three issues and only occasionally comment on possible patterns
of phase transitions.

\subsection{The orbit space}

The orbit space in 2HDM is not the entire Minkowski space. The square of the 4-vector $r^\mu$ is invariant under any
proper Lorentz transformation and, due to the Schwartz lemma, is non-negative
\be
r^2 \equiv r^\mu r_\mu = r_0^2 - r_i^2 = 4\left[(\fd_1\f_1)(\fd_2\f_2) - (\fd_1\f_2)(\fd_2\f_1)\right] \ge 0\,.
\ee
Therefore, the physical field configurations lie inside and on the border of the {\em future
light-cone ($LC^+$)} in the Minkowski space. The cone shape of the orbit space in 2HDM was known before,
see for example \cite{sartori2hdm}, however, its meaning in the context of Minkowski space was never realized.

The surface of $LC^+$, $r^2 = 0$, corresponds to the situation when the two Higgs doublets $\phi_1$ and $\phi_2$
are proportional to each other. In particular, if a vector $r^\mu$ indicates a stationary point (the vacuum) of the potential,
then $r^2 = 0$ means that the vacuum is electrically neutral. $r^2 > 0$ can be realized only when the two
doublets are not proportional; a vacuum solution with $r^2>0$ would correspond to the charge-breaking vacuum.

\subsection{Properties of $\Lambda^{\mu\nu}$}

The {\em positivity constraint} on the Higgs potential requires it to be
bounded from below. These constraints are usually presented as a list of inequalities among different
$\lambda$'s, see, for example, \cite{ginzreview}.
In our language the positivity constraint is given by a single statement:
\be
\mbox{{\em the tensor $\Lambda^{\mu\nu}$ is positive definite on the future light-cone}}\,,\label{positivity}
\ee
which, as is shown in Appendix~\ref{appsection-lambda}, is equivalent to the following
set of requirements:
\begin{itemize}
\item
tensor $\Lambda^{\mu\nu}$ is diagonalizable by an $SO(1,3)$ transformation,
\item
the timelike eigenvalue $\Lambda_0$ is positive,
\item
all spacelike eigenvalues $\Lambda_i$ are smaller than $\Lambda_0$.
\end{itemize}
In other words, there always exists an $SO(1,3)$ transformation that brings $\Lambda^{\mu\nu}$ to
$$
\left(\begin{array}{cccc}
\Lambda_0 & 0 & 0 & 0\\
0 & -\Lambda_1 & 0 & 0\\
0 & 0 & -\Lambda_2 & 0\\
0 & 0 & 0 & -\Lambda_3 \end{array}\right)\quad \mbox{with}\quad
\Lambda_0 > 0\ \mbox{and} \ \Lambda_0 > \Lambda_1, \Lambda_2, \Lambda_3\,.
$$
Note that $\Lambda_i$ are bounded only from above. The negative values of $\Lambda_i$
with arbitrary large absolute values are allowed. In particular, all $\Lambda_i < 0$ if and only if
$\Lambda^{\mu\nu}$ is positive definite in the entire Minkowski space.

Below we focus only on the physically relevant $\Lambda^{\mu\nu}$,
which satisfy (\ref{positivity}).

\subsection{The prototypical model}

Since $\Lambda^{\mu\nu}$ is diagonalizable, one can always perform such a $SO(1,3)$ rotation that
brings it to the diagonal form. By doing this, one arrives at the {\em prototypical model} of the initial
2HDM model. This model is defined by the diagonal $\Lambda^{\mu\nu}$, or explicitly, by
\be
\bar \lambda_1 = \bar \lambda_2 = \bar \lambda\,,\quad
\bar \lambda_6 = \bar \lambda_7 = 0\,, \quad  \Im\bar\lambda_5 = 0\,,\label{prototypical}
\ee
The potential of the prototypical model
\bea
V&=&-{1\over 2}\left[\bar{m}_{11}^2(\phi_1^\dagger\phi_1) + \bar{m}_{22}^2(\phi_2^\dagger\phi_2)
+ \bar{m}_{12}^2 (\phi_1^\dagger\phi_2) + \bar{m}_{12}^{2\ *} (\phi_2^\dagger\phi_1)\right]\label{Vproto}\\
&+& \fr{\bar\lambda}{2}\left[(\phi_1^\dagger\phi_1)^2 + (\phi_2^\dagger\phi_2)^2\right]
+\bar\lambda_3(\phi_1^\dagger\phi_1) (\phi_2^\dagger\phi_2)
+\bar\lambda_4(\phi_1^\dagger\phi_2) (\phi_2^\dagger\phi_1) +\fr{1}{2}\bar\lambda_5\left[(\phi_1^\dagger\phi_2)^2+
(\phi_2^\dagger\phi_1)^2\right]\nonumber
\eea
has only four free parameters in the quartic potential
$\bar\lambda,\,\bar\lambda_3,\,\bar\lambda_4,\,\bar\lambda_5$ and a generic set of the mass terms.

Thanks to Proposition~\ref{prop-positive-definite},
issues 1--4 listed in Section~\ref{sect-kinetic} depend on Lorentz-invariant quantities, and, therefore, can be studied at the level
of prototypical model. The answers to these questions are the same\footnote{There is a subtlety
with spontaneous $CP$-violation; see Section~\ref{sectionCP}} for the entire six-dimensional manifold
of 2HDM's that are linked to the given prototypical model by $SO(1,3)$ transformations\footnote{One can further reduce
the number of essential free parameters without losing the generality
by rescaling the fields and switching from $\bar\lambda,\,\bar\lambda_3,\,\bar\lambda_4,\,\bar\lambda_5$ to $1,\,\bar\lambda_3/\bar\lambda,
\,\bar\lambda_4/\bar\lambda,\,\bar\lambda_5/\bar\lambda$.}.
Therefore, if one intends to study issues 1--4 in the most general 2HDM, one can immediately
get rid off $\lambda_6$, $\lambda_7$, and $\Im\lambda_5$ and set $\lambda_1 = \lambda_2$,
without any lack of generality\footnote{In principle, the boost transformations do not remove the free parameters
but rather move them to the kinetic sector of the Higgs lagrangian. However, as was shown before,
the exact structure of the kinetic term has little relevance to the issues we study here. This comment is due to Roman Li.}.

We stress that we {\em do not} write down explicitly the expressions relating $\bar \lambda_i$
with the coupling constants $\lambda_i$ of the generic 2HDM. This would require
solving a fourth-order equation, and the expressions would be just impractical.
The point is that in order to understand the coarse-grained structure of a generic 2HDM
we {\em do not need} this exact transformation law.
We do not calculate nor use the numerical values of vacuum expectation values or masses of physical
Higgs bosons. If one is interested in properties of a 2HDM with particular values of coupling constants,
one can always diagonalize $\Lambda^{\mu\nu}$ numerically and immediately check what class of models
the given model belongs to.

\section{Charge-breaking vacuum}\label{section-charge}

\subsection{Stationary points}

Representation (\ref{Vmunu}) is a convenient starting point to study the properties
of the vacuum in 2HDM. The orbit space one obtains upon the spin map $\phi_i \to r^\mu$ is
a region with border ($LC^+$) in the Minkowski space.
Two possibilities must be considered in search for stationary points
(or better to say, {\em stationary orbits}) of the potential.
The first one is when a stationary orbit in the Higgs field space corresponds also to a stationary point
in the orbit space.
In a generic case it lies {\em inside} $LC^+$ and corresponds to the charge-breaking vacuum.
The second possibility is that a stationary orbit in the space of Higgs fields corresponds to a non-stationary point
on the {\em border} of the orbit space, i.e. on $LC^+$, which corresponds to a neutral vacuum.
In this Section we consider the former case, while Section 4 will be devoted to the latter one.

The stationary orbit is composed of stationary points, and
the conditions for the stationary point of the Higgs potential are
\be
{\partial V \over \partial \phi_i^\dagger} = {\partial r^\mu \over \partial \phi_i^\dagger} {\partial V \over \partial r^\mu}
= d_i^\mu \zeta_\mu = 0\,,\quad i = 1,2\,,\quad
\mbox{where} \quad d_i^\mu = \sigma^\mu_{ij} \phi_j\,,\quad \zeta^\mu = - M^\mu + \Lambda^{\mu\nu} r_\nu\,.\label{stationary}
\ee
Eq.~(\ref{stationary}) is a system of four equations since both real and imaginary parts are set to zero.
Introducing the light-cone vectors $n_+^\mu = (1,0,0,1)$, $n_-^\mu = (1,0,0,-1)$
and the ``transverse" unit vectors $e_1^\mu = (0,1,0,0)$, $e_2^\mu = (0,0,1,0)$, one can rewrite
(\ref{stationary}) as
\be
\phi_1 n_+^\mu \zeta_\mu + \phi_2 (e_1^\mu - i e_2^\mu )\zeta_\mu = 0\,,\quad
\phi_2 n_-^\mu \zeta_\mu + \phi_1 (e_1^\mu + i e_2^\mu )\zeta_\mu = 0\,.\label{stationarysep}
\ee
Consider the case of charge breaking vacuum. The two doublets are non-zero and are {\em not proportional}
to each other, so that each of (\ref{stationarysep}) splits into a pair of conditions that correspond
to vanishing coefficients in front of $\phi_1$ and $\phi_2$ separately.
One obtains
\be
\zeta_\mu n_+^\mu = 0\,,\quad
\zeta_\mu n_-^\mu = 0\,,\quad
\zeta_\mu e_1^\mu = 0\,,\quad
\zeta_\mu e_2^\mu = 0\,,\label{zeta}
\ee
from which one gets $\zeta^\mu = 0$, or
\be
\Lambda^{\mu\nu} r_\nu = M^\mu\,.\label{stationary2}
\ee
This is an inhomogeneous system of linear equations. In order to classify all situations here,
let us introduce the {\em image of $LC^+$} under the action of tensor $\Lambda^{\mu\nu}$:
\be
LC' = \{x^\mu | x^\mu = \Lambda^{\mu\nu}r_\nu \ \mbox{for all} \ r^\mu \in LC^+\}\,.
\ee
The following criterion holds for existence of the charge-breaking stationary point(s) in the orbit space:
\begin{proposition}\label{prop-charged-exist}
Charge-breaking stationary points exist if and only if $M^\mu$ lies inside the lightcone image $LC'$.
\end{proposition}
\begin{proof}
The proof follows from Eq.~(\ref{stationary2}) and from the properties of the operator $\Lambda^{\mu\nu}$
described in Appendix~\ref{appsection-lambda}.
Indeed, a point lies inside a cone if and only if the image of this point lies inside the image of the cone.
Since $r^\mu$ lies inside $LC^+$, $\Lambda^{\mu\nu}r_\nu$ lies inside $LC'$, and in order
for physical solutions of (\ref{stationary2}) to exist, $M^\mu$ must lie inside $LC'$.
\end{proof}
The number of solutions depends on whether $\Lambda^{\mu\nu}$ is invertible.
\begin{enumerate}
\item
If all eigenvalues of $\Lambda^{\mu\nu}$ are non-zero, then $\Lambda^{\mu\nu}$ is an invertible
operator. The solution of (\ref{stationary2}) exists, is unique and is given by
\be
r^\mu = (\Lambda^{-1})^{\mu} {}_\nu M^\nu \label{linearsolution}\,.
\ee
It corresponds to a physical solution only when $M^\nu$ lies inside $LC'$.
The value of the potential at this point is
$$
V = -{1\over 2}(\Lambda^{-1})^{\mu\nu} M_\mu M_\nu\,.
$$
In the language of the prototypical model (i.e. in the frame where $\Lambda^{\mu\nu}$ is diagonal),
one has
$$
V = -{1\over 2}\left({M_0^2 \over \Lambda_0} - \sum_i {M_i^2 \over \Lambda_i} \right)\,.
$$
Since $M^\mu$ lies inside $LC'$, this value is always negative.
\item
If at least one of $\Lambda_i$ is zero, then $\Lambda^{\mu\nu}$ is not invertible.
The kernel of $\Lambda^{\mu\nu}$, $\mathrm{Ker}\,\Lambda$, is a subspace
in the orbit space, and $LC'$ collapses to a manifold of smaller dimension. If $M^\mu \not \in LC'$,
there are no solutions. Otherwise,
one has a {\em continuum of charged solutions} whose dimension is equal to the dimension of $\mathrm{Ker}\,\Lambda$.
If one consider $\tilde{\Lambda}^{\mu\nu}$ and $\tilde M^\mu$, restrictions of $\Lambda^{\mu\nu}$ and $M^\mu$
to the space orthogonal to $\mathrm{Ker}\,\Lambda$, then $\tilde{\Lambda}^{\mu\nu}$ is invertible and a generic solution can be
represented as
\be
r^\mu = (\tilde{\Lambda}^{-1})^{\mu\nu} \tilde M_\nu + \rho^\mu\,,\quad\mbox{where}\quad \rho^\mu \in \mathrm{Ker}\,\Lambda\,.
\ee
The value of the potential at these points is
$$
V = -{1\over 2}(\tilde{\Lambda}^{-1})^{\mu\nu} {\tilde M}_\mu {\tilde M}_\nu\,.
$$
\end{enumerate}

\subsection{Mass matrix}

The stationary points found above can correspond either to the minima or the saddle points
of the potential. The non-trivial maxima cannot exist for a polynomial potential of fourth degree bounded from below.
The simplest proof is given by the directional minima technique discussed in
\cite{Kim}.
One first chooses some direction in the Higgs field space and considers a ray starting from the origin
along this direction. The Higgs potential depends on the coordinate $r$
along this ray as $V = \bar {M} r^2 + \bar \lambda r^4$
with some real $\bar {M}$ and some strictly positive $\bar \lambda$. The nontrivial stationary point
of this function can be only a minimum; this is the directional minimum.

Now one considers all possible directions in the Higgs field space and the manifold of all such directional minima.
The true stationary point of the potential must belong to this manifold; therefore at this stationary point
there always exists at least one direction along which the potential is concave upwards.
Therefore a non-trivial maximum of the potential,
which requires that the potential be strictly concave down at this point, is impossible.

In order to find the nature of each stationary point, one needs to calculate the second derivative matrix.
Upon differentiating in respect to the fields, the electroweak index ``opens up", and
one must trace it down carefully. The Higgs fields will be now written as $\phi_{i,\alpha}$, with $i=1,2$ labeling the first
and second doublet, while $\alpha =1,2$ labeling the up and down components of each doublet.
It is also convenient to switch explicitly to real and imaginary parts of the fields.
Since $\zeta^\mu = 0$ for the charged vacuum stationary points, one gets:
\bea
{\partial^2 V \over \partial \Re \phi_{i,\alpha}\, \partial \Re \phi_{j,\beta}} & = &
4\Re d^\mu_{i,\alpha} \Re d^\nu_{j,\beta}\,\Lambda_{\mu\nu}\,,\nonumber\\
{\partial^2 V \over \partial \Im \phi_{i,\alpha}\, \partial \Im \phi_{j,\beta}} & = &
4\Im d^\mu_{i,\alpha} \Im d^\nu_{j,\beta}\,\Lambda_{\mu\nu}\,,\nonumber\\
{\partial^2 V \over \partial \Re \phi_{i,\alpha}\, \partial \Im \phi_{j,\beta}} & = &
4\Re d^\mu_{i,\alpha} \Im d^\nu_{j,\beta}\,\Lambda_{\mu\nu}\,.\label{massCharged1}
\eea
Now, to merge these equations into one, consider the real vector $\varphi_{i,\alpha}$ with $i=1,\dots,4$ defined as:
\be
\varphi_{i,\alpha} = (\Re \phi_{1,\alpha},\, \Im \phi_{1,\alpha},\,
\Re \phi_{2,\alpha},\, \Im \phi_{2,\alpha})\,.\label{fieldsvar}
\ee
The second derivative matrices (\ref{massCharged1}) are then
\be
{\partial^2 V \over \partial \varphi_{i,\alpha}\, \partial \varphi_{j,\beta}} =
4 \Sigma_{ii'}^\mu \Sigma_{jj'}^\nu \Lambda_{\mu\nu}\cdot \varphi_{i',\alpha} \varphi_{j',\beta}\,,
\label{massCharged1.1}
\ee
where
\be
\Sigma_{ii'}^\mu = \left(\begin{array}{cccc}
n_+^\mu & 0 & e_1^\mu & e_2^\mu \\
0 & n_+^\mu & -e_2^\mu & e_1^\mu \\
e_1^\mu & -e_2^\mu & n_-^\mu & 0 \\
e_2^\mu & e_1^\mu & 0 & n_-^\mu
\end{array}\right)\,.\label{bigsigma}
\ee
As a final step, we merge all $\varphi_{i,\alpha}$ into a single 8D real vector
\be
\varphi_a = (\varphi_{i,1},\, \varphi_{i,2})\,,\quad a = 1,\, \dots,\, 8\,,\label{fieldsvar2}
\ee
and for any pair of indices $a=\{i,\,\alpha\}$ and $a'=\{i',\,\alpha'\}$
define
\be
\Sigma_{aa'}^\mu = \Sigma_{ii'}^\mu \cdot \delta_{\alpha\alpha'}\,.\label{bigsigma2}
\ee
Then the second derivative matrix takes form
\be
{\partial^2 V \over \partial \varphi_{a}\, \partial \varphi_{b}} =
4 \Sigma_{aa'}^\mu \Sigma_{bb'}^\nu \Lambda_{\mu\nu}\cdot \varphi_{a'} \varphi_{b'}\,,
\label{massCharged1.2}
\ee

Note that $\Sigma_{aa'}^\mu$ is invariant under rotations that mix
the real and imaginary parts of the same field or mix the electroweak indices $\alpha$ and $\beta$.
This leads to the observation that matrix (\ref{massCharged1.2})
is {\em covariant} under electroweak transformations, which is, of course, a plain consequence
of the scalar potential $V(\phi)$ being an EW-scalar.
This fact means that the second derivative matrix must have four
zero modes, which correspond to simultaneous EW-rotations inside doublets $\phi_1$ and $\phi_2$.

In order to study the positive-definiteness of the matrix (\ref{massCharged1.1}),
we exploit Proposition~\ref{prop-positive-definite} and perform a convenient $SO(1,3)$ rotation.
Since the charged vacuum solution $r^\mu$ lies inside the $LC^+$,
one can always perform such a $SO(1,3)$ rotation that makes $r^\mu
= (u^2,\, 0,\, 0,\, 0)$, with real $u$. Then, one is free to
choose a representing point on this orbit by performing a
convenient EW rotation. A particularly simple representing point
is
$$
\phi_1 = {1\over \sqrt{2}}\stolbik{0}{u}\,,\quad \phi_2 = {1\over \sqrt{2}}\stolbik{u}{0}\,.
$$
The second derivative matrix (\ref{massCharged1.2}) written in this reference frame is
\be
{\partial^2 V \over \partial \varphi_{a}\, \partial \varphi_{b}} =
2 u^2 \left(\begin{array}{cccccccc}
\bracket{e_1}{e_1} & - \bracket{e_1}{e_2} & \bracket{e_1}{n_-} & 0 & \bracket{e_1}{n_+} & 0 & \bracket{e_1}{e_1} & \bracket{e_1}{e_2} \\
-\bracket{e_2}{e_1} & \bracket{e_2}{e_2} & -\bracket{e_2}{n_-} & 0 &-\bracket{e_2}{n_+} & 0 &-\bracket{e_2}{e_1} &-\bracket{e_2}{e_2} \\
\bracket{n_-}{e_1} & - \bracket{n_-}{e_2} & \bracket{n_-}{n_-} & 0 & \bracket{n_-}{n_+} & 0 & \bracket{n_-}{e_1} & \bracket{n_-}{e_2} \\
0 & 0 & 0 & 0 & 0 & 0 & 0 & 0 \\
\bracket{n_+}{e_1} & - \bracket{n_+}{e_2} & \bracket{n_+}{n_-} & 0 & \bracket{n_+}{n_+} & 0 & \bracket{n_+}{e_1} & \bracket{n_+}{e_2} \\
0 & 0 & 0 & 0 & 0 & 0 & 0 & 0 \\
\bracket{e_1}{e_1} & - \bracket{e_1}{e_2} & \bracket{e_1}{n_-} & 0 & \bracket{e_1}{n_+} & 0 & \bracket{e_1}{e_1} & \bracket{e_1}{e_2} \\
\bracket{e_2}{e_1} & - \bracket{e_2}{e_2} & \bracket{e_2}{n_-} & 0 & \bracket{e_2}{n_+} & 0 & \bracket{e_2}{e_1} & \bracket{e_2}{e_2}
\end{array}\right)\,,
\ee
where we introduced short notation
\be
\bracket{p}{q} \equiv p^\mu q^\nu \Lambda_{\mu\nu}\,.\label{bracket}
\ee
Upon obvious rotations inside four pair of coordinates, one arrives
at the above mentioned four Goldstone modes, which in this particular basis correspond to
$$
{1\over\sqrt{2}}(\Im\phi_{1,2} \pm \Im\phi_{2,1})\,,\quad
{1\over\sqrt{2}}(\Im\phi_{1,1} + \Im\phi_{2,2})\,,\quad
{1\over\sqrt{2}}(\Re\phi_{1,1} - \Re\phi_{2,2})\,.
$$
When gauge sector is taken into account, these modes will become the longitudinal components
of the four gauge bosons, including the photon. The other four modes
have the following second derivative matrix
\be
M_{ij}^2 \equiv {1\over 2} {\partial^2 V \over \partial \varphi_i\partial \varphi_j} = 2 u^2 \left(
\begin{array}{cccc}
\bracket{e_0}{e_0} & \bracket{e_0}{e_1} & \bracket{e_0}{e_2} & \bracket{e_0}{e_3} \\
\bracket{e_1}{e_0} & \bracket{e_1}{e_1} & \bracket{e_1}{e_2} & \bracket{e_1}{e_3} \\
\bracket{e_2}{e_0} & \bracket{e_2}{e_1} & \bracket{e_2}{e_2} & \bracket{e_2}{e_3} \\
\bracket{e_3}{e_0} & \bracket{e_3}{e_1} & \bracket{e_3}{e_2} & \bracket{e_3}{e_3}
\end{array}
\right)\,,\label{massCharged3}
\ee
with
$$
e_0^\mu = {1\over 2}(n_+^\mu + n_-^\mu) = (1,\,0,\,0,\,0)\,,
\quad e_3^\mu = {1\over 2}(n_+^\mu - n_-^\mu) = (0,\,0,\,0,\,1)\,,
$$
and $e_1^\mu$, $e_2^\mu$ defined at the beginning of this section.
We remind that the matrix (\ref{massCharged3}) is not the mass matrix of the physical Higgs bosons,
see discussion in Sect.~\ref{sect-kinetic}, but is linked to it by a certain $SL(4,R)$ transformation.

Despite the very suggestive form of (\ref{massCharged3}), there is no simple relation between
the sets of eigenvalues of $\Lambda^{\mu\nu}$ and the mass matrix.
Indeed, the matrix in (\ref{massCharged3}) has form
$$
\left(
\begin{array}{cccc}
\Lambda_{00} & - \Lambda_{0i} \\
- \Lambda_{0j} & \Lambda_{ij}
\end{array}
\right)\,,
$$
while the matrix $\Lambda^\mu{}_\nu = \Lambda^{\mu\nu'} g_{\nu'\nu}$ is
$$
\left(
\begin{array}{cccc}
\Lambda_{00} & - \Lambda_{0i} \\
\Lambda_{0j} & - \Lambda_{ij}
\end{array}
\right)\,.
$$
However, it is easy to establish the following general condition for the charge-breaking minimum:
\begin{proposition}\label{prop-charged}
The charge-breaking stationary point is a minimum if and only if $\Lambda^{\mu\nu}$ is positive definite
in the entire Minkowski space.
\end{proposition}
\begin{proof}
Suppose $p_i = (p_0,\ p_1,\ p_2,\ p_3)$ is a normalized eigenvector of (\ref{massCharged3}) with the eigenvalue $M^2$.
Then
$$
M^2 = M_{ij}^2 p_i p_j = u^2 \Lambda^{\mu\nu} P_\mu P_\nu\,,\quad \mbox{where} \quad
P^\mu \equiv p_0 e_0^\mu + p_1 e_1^\mu + p_2 e_2^\mu + p_3 e_3^\mu\,.
$$
If $\Lambda^{\mu\nu}$ is positive definite in the entire space, $\Lambda^{\mu\nu} P_\mu P_\nu > 0 \ \forall P_\mu$, and
all the eigenvalues of the mass matrix are positive. The stationary point is, therefore, a minimum,
and according to Proposition~\ref{prop-positive-definite}, it remains a minimum
when transformed back to the ``preferred frame" with canonical kinetic term.

Inversely, if $\Lambda^{\mu\nu}$ is not positive definite, then a vector $Q_\mu$ exists
such that $\Lambda^{\mu\nu} Q_\mu Q_\nu < 0$. Since $\{e_0^\mu,\, e_1^\mu,\, e_2^\mu,\, e_3^\mu\}$
form a basis in the Minkowski space, $Q_\mu$ can be uniquely represented as
$Q^\mu = q_0 e_0^\mu + q_1 e_1^\mu + q_2 e_2^\mu + q_3 e_3^\mu$.
Then,
$$
u^2 \Lambda^{\mu\nu} Q_\mu Q_\nu = M_{ij}^2 q_i q_j.
$$
Representing $q_i$ in the basis of eigenvectors $\{p^{(a)}_i\}$ of the mass matrix,
$q_i = \sum_a c_{(a)} p_i^{(a)}$, one simplifies this to
$$
\sum_a M_{a}^2 c_{(a)}^2 <0\,,
$$
which can be satisfied only when at least one of mass matrix eigenvalues is negative.
\end{proof}
This Proposition might be also guessed from the expression for the second derivative matrix (\ref{massCharged1.2}).

Thus, in order for the charge-breaking stationary point to be a minimum, all $\Lambda_i$ must be non-positive.
In the language of the prototypical model, the charge-breaking minimum takes place, when
\be
\bar\lambda_3 < \bar\lambda\,,\quad \bar\lambda_4 >0\,,\quad |\bar\lambda_5| < \bar\lambda_4\,.
\ee
Note, finally, that if $\Lambda^{\mu\nu}$ has at least one trivial eigenvalue, then the mass matrix will
also have trivial eigenvalues, i.e. additional Goldstone modes. This can be most easily
seen from the fact that $\det\Lambda^{\mu}_{\cdot}{}_\nu = - \det M^2_{ij}$.

\section{Neutral vacuum}

\subsection{Strategy for finding solutions and general expression for the mass matrix}\label{section-strategy}

The neutral vacuum solutions are located on the border of the orbit space. Therefore
a neutral stationary orbit of the potential in the space of Higgs fields does not necessarily
correspond to a stationary point in the orbit space.
However, in order to keep the terminology simple, we will use term ``neutral stationary point" for
``point in the orbit space that corresponds to the neutral stationary orbit in the space of Higgs fields".

Since $LC^+$ is invariant under $SO(1,3)$, one simplifies the analysis from the start by
making an appropriate boost to align the ``time-like" principal axis
of $\Lambda^{\mu\nu}$ with the future line, so that all $\Lambda^{0i} = 0$.
After this, one will need to perform only 3D rotations.

For the neutral vacuum solution, the two Higgs doublets are {\em proportional}, and equations (\ref{stationary})
are not reduced anymore into four equations (\ref{zeta}).
In order to find all the solutions, we use the following strategy. Take a point
$r^\mu = {1\over 2}v^2(1,\ \vec{n})$ lying on $LC^+$ and perform a 3D rotation that makes $r^\mu = {1\over 2}v^2 n_+^\mu$,
which corresponds to setting $\phi_2 = 0$. In this way one obtains three instead of four equations
(\ref{zeta}); the only difference is that now $\zeta_\mu n_-^\mu \not = 0 \to \zeta^\mu = \zeta n_+^\mu$.
The solution of the resulting equation
\be
{1\over 2}\Lambda^{\mu\nu} \cdot v^2 n_{+\, \nu} - \zeta n_+^\mu = M^\mu\,.\label{stationary3}
\ee
exists, and is unique, if $M^\mu$ lies on the half plane
$\Lambda^{\mu\nu} \cdot v^2 n_{+\, \nu}/2 - \zeta n_+^\mu$, parameterized by real numbers $v^2 > 0$ and $\zeta$.
Note that this plane never collapses
to a line, because $\Lambda^{\mu\nu}$ does not have any non-trivial eigenvectors $r^\mu$ on the $LC^+$.
The values of $v^2$ and $\zeta$ in this Lorentz frame are
\be
v^2 = {2 M_\mu n_+^\mu \over \langle n_+ | n_+\rangle}\,,\quad
\zeta = -{1\over 2}M_\mu n_-^\mu + {1\over 4}v^2 \langle n_+ | n_-\rangle\,,\label{v2zeta}
\ee
where we used short notation (\ref{bracket}).
Since $n^\mu_+ + n^\mu_- = 2e^\mu_0$ and due to $\Lambda^{0i} = 0$, one can simplify $\zeta$ to
\be
\zeta = -{1\over 2}M_\mu n_-^\mu + {1 \over 2}v^2 \Lambda_0 - {1\over 4}v^2 \langle n_+ | n_+\rangle =
{1 \over 2}v^2 \Lambda_0 - M_0\,,
\label{zeta2}
\ee
so that if one fixes $M_0$, the variables $\zeta$ and $v^2$ are linearly dependent.
The depth of the potential at the stationary point is
\be
V = -{1\over 4}v^2M_\mu n_+^\mu = -{1\over 2} {(M_\mu n_+^\mu)^2 \over \langle n_+ | n_+\rangle}\,.
\ee
In contrast to the values of $v^2$ and $\zeta$, this depth is a Lorentz-independent quantity.

Repeat now the same check for {\em every possible} lightcone vector.
Geometrically, we take every point $r^\mu = {1 \over 2}v^2(1, \vec n)$, find its image $(r')^\mu$
under the action of $\Lambda^{\mu\nu}$, attach a line generated by $r^\mu$ at this point,
and check if $M^\mu$ lies on it. As we continuously check all possible lightcone vectors,
the half-plane generated by $v^2$ and $\zeta$ sweeps some region in the orbit space.
Therefore, the problem of finding all stationary point of the Higgs potential is
reformulated as the problem of finding this region and establishing how many times each point
is swept as we check all possible vectors $r^\mu \in LC^{+}$.

This problem will be solved below in geometric terms, see Proposition~\ref{prop-number}. But one fact is plain to see now:
\begin{proposition}\label{prop-exist}
Non-trivial solutions to the stationary point problem can exist only if $M^\mu$ lies outside the past light-cone $LC^-$.
\end{proposition}
\begin{proof}
The representation of $v^2$ in (\ref{v2zeta}) requires that $M_\mu n^\mu > 0$ for some lightcone vector $n^\mu$.
But if $M^\mu$ lies in $LC^-$, then for any vector $n^\mu \in LC^+$ the scalar product $M_\mu n^\mu$ is negative
(this becomes obvious after making a boost to align $M^\mu$ with the past direction).
Therefore, if $M^\mu$ lies inside the past light-cone $LC^-$, there are no nontrivial solutions.
\end{proof}

The second derivative matrix at a generic neutral stationary point can be also found.
In the case of the neutral vacuum solution, $\zeta^\mu \not = 0$.
Therefore, in the second derivative of the potential one needs to differentiate not only $\zeta^\mu$, but also $d_i^\mu$.
One obtains
\bea
{\partial^2 V \over \partial \Re \phi_{i,\alpha}\, \partial \Re \phi_{j,\beta}} & = &
2\Re (\sigma^\mu_{ij} \zeta_\mu)\,\delta_{\alpha\beta} +
4\Re d^\mu_{i,\alpha} \Re d^\nu_{j,\beta}\,\Lambda_{\mu\nu}\,,\nonumber\\
{\partial^2 V \over \partial \Im \phi_{i,\alpha}\, \partial \Im \phi_{j,\beta}} & = &
2\Re (\sigma^\mu_{ij} \zeta_\mu)\,\delta_{\alpha\beta} +
4\Im d^\mu_{i,\alpha} \Im d^\nu_{j,\beta}\,\Lambda_{\mu\nu}\,,\nonumber\\
{\partial^2 V \over \partial \Re \phi_{i,\alpha}\, \partial \Im \phi_{j,\beta}} & = &
- 2\Im (\sigma^\mu_{ij} \zeta_\mu)\,\delta_{\alpha\beta} +
4\Re d^\mu_{i,\alpha} \Im d^\nu_{j,\beta}\,\Lambda_{\mu\nu}\,.\label{massNeutral1}
\eea
Again, one can merge these matrices into a single expression using fields $\varphi_{a}$, $a=1,\,\dots,\, 8$
defined in (\ref{fieldsvar2}) and $\Sigma_{aa'}^\mu$ defined in (\ref{bigsigma2}):
\be
{\partial^2 V \over \partial \varphi_{a}\, \partial \varphi_{b}} =
2 \Sigma_{ab}^\mu \,\zeta_\mu
+ 4 \Sigma_{aa'}^\mu \Sigma_{bb'}^\nu \Lambda_{\mu\nu}\cdot \varphi_{a'} \varphi_{b'}\,,
\label{massNeutral1.1}
\ee
To simplify the analysis, perform a 3D rotation\footnote{Note that $\Lambda^{ij}$, in general, is not diagonalized
by this rotation.}
to make $r^\mu = {1\over 2}v^2 n_+^\mu$ and $\zeta^\mu = \zeta n_+^\mu$,
and then an EW rotation to choose the representative point
$$
\phi_1 = {1\over\sqrt{2}}\stolbik{0}{v}\,,\quad \phi_2 = \stolbik{0}{0}\,,
$$
of the orbit.
This choice leads to the following expression for the second derivative matrix
\be
{\partial^2 V \over \partial \varphi_{a}\, \partial \varphi_{b}} =
\left(\begin{array}{cccccccc}
0 & 0 & 0 & 0 & 0 & 0 & 0 & 0 \\
0 & 0 & 0 & 0 & 0 & 0 & 0 & 0 \\
0 & 0 & 4\zeta & 0 & 0 & 0 & 0 & 0 \\
0 & 0 & 0 & 4\zeta & 0 & 0 & 0 & 0 \\
0 & 0 & 0 & 0 & 2v^2 \bracket{n_+}{n_+} & 0 & 2v^2 \bracket{n_+}{e_1} & 2v^2 \bracket{n_+}{e_2} \\
0 & 0 & 0 & 0 & 0 & 0 & 0 & 0 \\
0 & 0 & 0 & 0 & 2v^2 \bracket{e_1}{n_+} & 0 & 4\zeta + 2v^2 \bracket{e_1}{e_1} & 2v^2 \bracket{e_1}{e_2} \\
0 & 0 & 0 & 0 & 2v^2 \bracket{e_2}{n_+} & 0 & 2v^2 \bracket{e_2}{e_1} & 4\zeta + 2v^2 \bracket{e_2}{e_2}
\end{array}\right)\,,
\ee
One sees that the modes corresponding to charged ($\alpha = \beta = 1$) and neutral
($\alpha = \beta = 2$) excitations split. For the charged modes, one has two Goldstones
and two massive modes with eigenvalues $4\zeta$.
Among the four neutral modes there is one Goldstone, $\Im\phi_{1,2}$, while the other three modes,
$\varphi_i \equiv (\Re\phi_{1,2},\, \Re\phi_{2,2},\, \Im\phi_{2,2})$, have the following mass matrix
\be
M_{ij} = {1\over 2}{\partial^2 V \over \partial \varphi_i \partial \varphi_j}  =
\left(\begin{array}{ccc}
0 & 0 & 0 \\
0 & 2\zeta & 0 \\
0 & 0 & 2\zeta
\end{array}\right)
+ v^2 \left(\begin{array}{ccc}
\bracket{n_+}{n_+} & \bracket{n_+}{e_1} & \bracket{n_+}{e_2} \\
\bracket{e_1}{n_+} & \bracket{e_1}{e_1} & \bracket{e_1}{e_2} \\
\bracket{e_2}{n_+} & \bracket{e_2}{e_1} & \bracket{e_2}{e_2}
\end{array}\right)\,.\label{massNeutral3}
\ee
Since $\bracket{n_+}{n_+}$ is positive, there is always a positive eigenvalue of $M_{ij}$.
With $\Lambda^{0i}=0$ and (\ref{zeta2}), the mass matrix simplifies to
\be
M_{ij}  =
v^2 \left(\begin{array}{rrr}
\Lambda_0+\bracket{e_3}{e_3} & \bracket{e_3}{e_1} & \bracket{e_3}{e_2} \\
\bracket{e_1}{e_3} & \Lambda_0+\bracket{e_1}{e_1} & \bracket{e_1}{e_2} \\
\bracket{e_2}{e_3} & \bracket{e_2}{e_1} & \Lambda_0+\bracket{e_2}{e_2}
\end{array}\right)
+
\left(\begin{array}{ccc}
0 & 0 & 0 \\
0 & - 2M_0 & 0 \\
0 & 0 & - 2M_0
\end{array}\right)\,.\label{massNeutral4}
\ee
In order for the stationary point to be minimum, one must simultaneously have $\zeta > 0$ and all
eigenvalues of (\ref{massNeutral4}) positive. Although the eigenvalues of both matrices in (\ref{massNeutral4})
are known, $v^2(\Lambda_0-\Lambda_i)>0$ for the first matrix
and $0$, $-2M_0$ for the second matrix, the eigenvalues of their sum are not related
in a simple way to them, since these matrices do not commute.
The explicit expressions for the eigenvalues
can be written down, but they do not give any simple minimum criterion in the general case.
However, when $\Lambda^{\mu\nu}$ is positive definite
in the entire Minkowski space, the analysis simplifies and the following important fact can be easily proved:
\begin{proposition}\label{prop-coexist}
The charge-breaking and neutral minima never coexist in 2HDM.
\end{proposition}

\begin{proof}
According to Proposition~\ref{prop-charged-exist}, a charge-breaking solution exists if and only if $M^\mu$
lies inside lightcone image $LC'$.
Besides, as was proved in Proposition~\ref{prop-charged}, the charge-breaking stationary point is a minimum if and only if $\Lambda^{\mu\nu}$
is positive definite in the entire Minkowski space, i.e. when all $\Lambda_i < 0$.

If $\Lambda^{\mu\nu}$ is positive definite in the entire Minkowski space,
the second matrix in (\ref{massNeutral3}) becomes positive definite, too. Therefore,
in this case the neutral stationary point is a minimum if and only if $\zeta>0$.
But if $\zeta>0$, then $M^\mu$ lies outside the image of the lightcone $LC'$. Indeed, rewriting (\ref{stationary3})
with the lightcone vector $n^\mu = (1,\,n_1,\,n_2,\,n_3)$
in the coordinate frame where $\Lambda^{\mu\nu}$ is diagonal, one gets
$$
{1\over 2}v^2 (\Lambda_0,\,-|\Lambda_1|n_1,\,-|\Lambda_2|n_2,\,-|\Lambda_3|n_3) - \zeta
(1,\,n_1,\,n_2,\,n_3) = M^\mu\,.
$$
In plain words, to arrive at $M^\mu$ starting from a point $r^{\prime\, \mu}$ on $LC'$,
one must decrease its timelike coordinate $r^\prime_0$, while
increasing the absolute values of {\em all} spacelike coordinates $r^\prime_i$. Such a shift brings us {\em outside}
$LC'$. Therefore, conditions for a charge-breaking minimum and neutral minimum are opposite and can never be met
simultaneously.
\end{proof}
In other words, if one has a charge-breaking minimum, then the neutral stationary points are necessarily saddle points,
and vice versa. This fact offers a very clear explanation to the finding
that if a neutral minimum exists, the charge-breaking stationary points lie above it, see \cite{barroso2006}
and references therein. Note also that Proposition~\ref{prop-coexist} implies that the charge-breaking or charge-restoring
phase transitions cannot be of the first order\footnote{Here we mean the zero-temperature phase transitions
from one minimum to another under the slow variation of the parameters of the Higgs potential. We do not
consider here thermal phase transitions.}.

Another observation can be read off Eq.~(\ref{massNeutral4}):
\begin{proposition}\label{prop-negative}
If the prototypical model has $M_0 \le 0$, then the neutral stationary point,
when exists, is always a minimum.
\end{proposition}
\begin{proof}
In deriving the mass matrix (\ref{massNeutral4}), we have already
aligned the timelike axis of $\Lambda^{\mu\nu}$ with the future direction.
To arrive at the prototypical model, one would need only a 3D rotation, which does not change $M_0$.
Therefore, the value of $M_0$ that enters (\ref{massNeutral4}) is the same as in the prototypical model.

If $M_0 \le 0$, then $\zeta = {1\over 2}v^2 \Lambda_0 - M_0 > 0$,
so that one needs to check only the eigenvalues of (\ref{massNeutral4}).
But this matrix is a sum of a strictly positive and a non-negative matrices and, therefore,
it is again positive definite.
Its eigenvalues are all positive, and the stationary point is a minimum.
\end{proof}

\subsection{Toy model}\label{toysect}

The analysis of the number and properties of the solutions in the general case will be given
later in geometric terms. Before we turn to it, it is instructive to consider in detail a toy model,
which introduces these geometric constructions
in a simple way.

The toy model is a 2HDM potential with three equal eigenvalues of $\Lambda^{\mu\nu}$:
$\Lambda_1=\Lambda_2=\Lambda_3=\Lambda$. Its prototypical model has $\bar \lambda_5 = 0$ and
$\bar \lambda = \bar \lambda_3 + \bar \lambda_4$ and is invariant under any 3D rotation.
We reiterate the point that the results obtained below
hold not only for the prototypical model itself, but also for all the 2HDM's related to it
by $SO(1,3)$ transformations.

The operator $\Lambda^{\mu\nu}$ in the toy model acting on the orbit space vectors preserves
the 3D spherical symmetry. Consider the plane in the orbit space defined by the future 4-vector $e_0^\mu = (1,0,0,0)$
and the mass terms 4-vector\footnote{If $M^\mu \propto e_0^\mu$,
there would be a whole sphere of solution to the stationary point problem.}
$M^\mu$. The stationary points $r^\mu$, if exist, must also lie on this plane.
Therefore, following the strategy for the search of the stationary points outlined above,
one must check only the two directions on the lightcone that belong also to this plane.

\begin{figure}[!htb]
   \centering
\includegraphics[width=12cm]{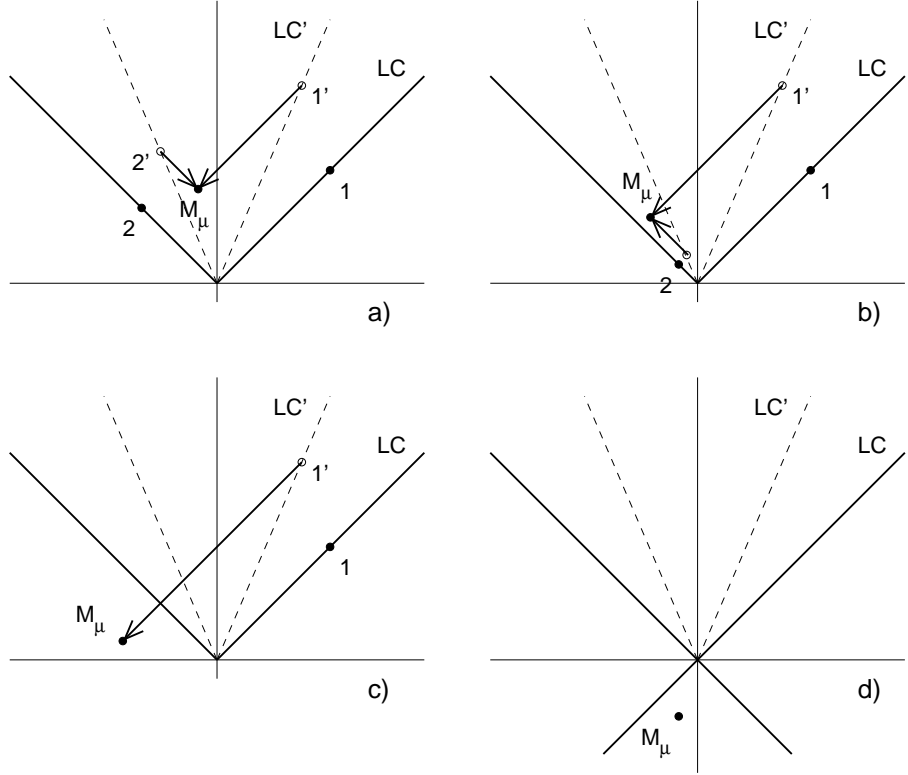}
\caption{Schematic representation of the search for the neutral stationary points for the toy model
with $\Lambda>0$. One takes a point $r^\mu$ on $LC^+$, finds its image $r^{\mu\,\prime}$ on $LC'$
under action of $\Lambda^{\mu\nu}$ and then subtracts $\zeta n^\mu$ to arrive at $M^\mu$. Arrows going down correspond to
$\zeta > 0$, arrows going up correspond to $\zeta<0$. Shown are the four possible positions of $M^\mu$:
(a) $M^\mu$ inside $LC'$, $\zeta > 0$ for both solutions; (b) $M^\mu$ inside $LC$ but outside $LC'$, $\zeta>0$ only for solution 1;
(c) $M^\mu$ outside $LC$ ($M^2<0$), the only solution with $\zeta >0$; (d) $M^\mu$ inside $LC^-$, no solutions.
}
   \label{fig-toy1}
\end{figure}

\begin{figure}[!htb]
   \centering
\includegraphics[width=12cm]{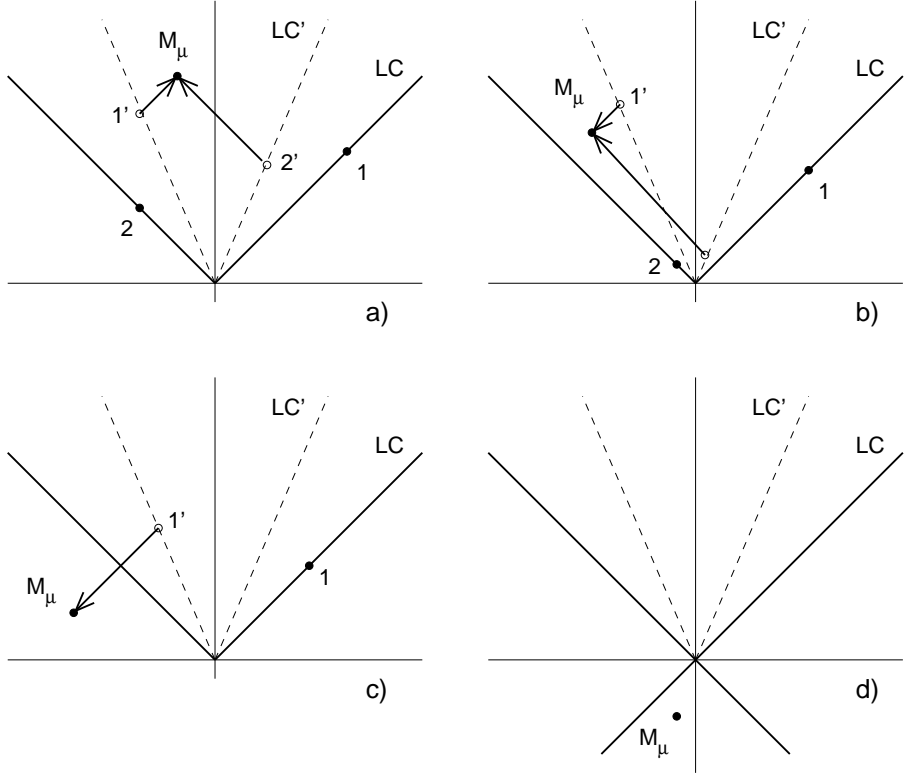}
\caption{The same as in Fig.~\ref{fig-toy1} but for $\Lambda<0$, $|\Lambda|<\Lambda_0$.
(a) $M^\mu$ inside $LC'$, $\zeta < 0$ for both solutions; (b) $M^\mu$ inside $LC$ but outside $LC'$, $\zeta>0$ only for solution 1;
(c) $M^\mu$ outside $LC$ ($M^2<0$), the only solution with $\zeta >0$; (d) $M^\mu$ inside $LC^-$, no solutions.
}
   \label{fig-toy2}
\end{figure}

The number of stationary points depends on the position of $M^\mu$, which is illustrated and explained in Figs.~\ref{fig-toy1}
($\Lambda>0$) and \ref{fig-toy2} ($\Lambda<0$).
\begin{itemize}
\item
If $M^\mu$ lies inside the future lightcone $LC^+$, then there are two neutral solutions.
In terms of the components of $M^\mu$ in this particular basis, this occurs when $\bar{m}_{11}^2,\,\bar{m}_{22}^2  > 0$.
\item
If $M^\mu$ lies outside $LC^+$ and outside the past lightcone $LC^-$ (i.e. $M^2 < 0$
implying $\bar{m}_{11}^2$ and $\bar{m}_{22}^2$ are of the opposite signs), then there is
a unique solution.
\item
If $M^\mu$ lies inside the past lightcone $LC^-$ ($\bar{m}_{11}^2,\,\bar{m}_{22}^2  < 0$), there are no solutions,
Fig.~\ref{fig-toy1}d.
\end{itemize}

The saddle point/minimum classification of these points depends on the following three situations:
\begin{itemize}
\item
$\Lambda>0$, which means $\bar\lambda_4 < 0$ and $\bar\lambda_3 = \bar\lambda - \bar\lambda_4 > \bar\lambda$;
\item
$\Lambda<0$ and $|\Lambda|<\Lambda_0$, which means $\bar\lambda > \bar\lambda_3 > 0$
and $\bar\lambda > \bar\lambda_4 > 0$;
\item
$\Lambda<0$ but $|\Lambda|>\Lambda_0$, which means $\bar\lambda_4 > \bar\lambda$ and $\bar\lambda_3 < 0$.
\end{itemize}
Let us consider the first case, shown in Fig.~\ref{fig-toy1}.
If $M^\mu$ lies inside $LC^+$, we have two solutions, one on each of the two light-rays. Without lack of generality,
we set $\bar{m}_{11}^2 > \bar{m}_{22}^2 > 0$ (i.e. $M_3 < 0$). In the chosen basis,
\bea
&&M_\mu n_+^\mu = M_0 - M_3 = {\bar{m}_{11}^2 \over 2}\,,\quad
M_\mu n_-^\mu = M_0 + M_3 = {\bar{m}_{22}^2 \over 2}\,,\nonumber \\
&&\bracket{n_+}{n_+} = \Lambda_0 - \Lambda = \bar\lambda\,,\quad
\bracket{n_+}{n_-} = \Lambda_0 + \Lambda = \bar\lambda_3\,,\quad
\bracket{e_2}{e_2} = - \Lambda\,.\nonumber
\eea
The eigenvalues of the matrix (\ref{massNeutral4}) simplify to
$v^2(\Lambda_0 - \Lambda)$ and twice degenerate value $v^2(\Lambda_0 - \Lambda) - 2M_0 = 2 \zeta - v^2 \Lambda$.
The two solutions are
\bea
\mbox{solution 1} && v^2 = 2{M_0 - M_3 \over \Lambda_0 - \Lambda}\,,\quad
V_1 = - 2{(M_0 - M_3)^2\over \Lambda_0 - \Lambda}\,,\nonumber\\
\mbox{solution 2} && v^2 = 2{M_0 + M_3 \over \Lambda_0 - \Lambda}\,,\quad V_2 = - 2{(M_0 + M_3)^2\over \Lambda_0 - \Lambda}\,.
\eea
The first solution lies deeper than the second; in fact, the first solution is a minimum, since
\be
2 \zeta =  2{\Lambda M_0 - \Lambda_0 M_3 \over \Lambda_0 - \Lambda} > 0
\quad \mbox{and} \quad 2 \zeta - v^2 \Lambda = - 2 M_3 >0 \,,
\ee
while the second solution is a saddle point due to $2 \zeta - v^2 \Lambda = 2M_3 < 0$.

If $M^\mu$ lies outside $LC^+$, then $\bar{m}_{22}^2$ becomes negative, and the second solution disappears.
The first solution, however, still corresponds to the minimum.

Let us turn to the case $\Lambda < 0$, which means that $\Lambda^{\mu\nu}$ is positive definite in the entire
Minkowski space. If $|\Lambda| < \Lambda_0$, then $LC'$ still lies inside $LC^+$ (this case is shown in Fig.~\ref{fig-toy2}),
while for $|\Lambda| > \Lambda_0$ the cone $LC'$ is wider than $LC^+$.

The above expressions are still valid in this case,
however the change of sign of $\Lambda$ changes also geometric constructions.
When $M^\mu$ lies both inside $LC^+$ and $LC'$, then,
as before, two solutions exist, but both of them have $\zeta < 0$, therefore,
both correspond to the saddle points. In this case the charge-breaking stationary point is the only minimum
of the potential, in agreement with Proposition~\ref{prop-coexist}.

For a neutral stationary point to become a minimum, $\zeta$ must be positive, which
happens when $M^\mu$ lies outside the lightcone image $LC'$.
Note that in the case the other neutral stationary point, if exists,
must be a saddle point, since the mass matrices for these two points
have among their eigenvalues $2M_3$ or $-2M_3$.

It is interesting to trace how the properties of the vacuum change if $\Lambda<0$ is fixed
and $M^\mu$ continuously shifts from inside to outside of the $LC'$. Initially the ground state
of the potential is charge-breaking, but upon crossing $LC'$ it turns into a neutral one.
The solution of the stationary point problem also moves continuously, and one observes a
charge-restoring phase transition of the second order. We stress that this is a zero-temperature phase transition,
not a thermal one.
Such an evolution of $M^\mu$, and therefore such a phase transition, might be relevant
for the early history of the Universe, \cite{ginzuniverse}.

We summarize the solution to the minimization problem in the toy model:
\begin{itemize}
\item
Non-trivial stationary points of the Higgs potential exist
if and only if $M^\mu$ lies outside the past lightcone, in agreement with Proposition~\ref{prop-exist}.
Among these stationary points, there is only one minimum.
If other stationary points exist, they are saddle points.
\item
If $\Lambda>0$, the minimum of the Higgs potential is neutral.
Depending on the position of $M^\mu$, it can be also
accompanied by one neutral and one charge-breaking saddle points.
\item
If $\Lambda <0$, then the minimum can be either neutral or charge-breaking.
If $M^\mu$ lies inside $LC'$, the minimum is charge-breaking, while for $M^\mu$
outside $LC'$, it is neutral.
\end{itemize}

\subsection{Geometric constructions in the toy model}

The above criteria for the existence and the number of solutions in the toy model
can be obtained in a purely geometric fashion.

As before, make appropriate boosts to set all $\Lambda_{0i} =0$ and
consider the following construction. Take a sphere composed of points
$r^\mu = {1 \over 2}v^2(1,\vec n)$ with fixed $v^2$ and all possible orientations of the unit 3D vector $\vec n$.
Under the action of $\Lambda^{\mu\nu}$ it is mapped to another sphere
$r^{\prime\mu} = {1 \over 2}v^2(\Lambda_0,\Lambda \vec n)$. At each point of this sphere attach now a straight
line parallel to $r^\mu$, i.e. consider the 3D manifold, defined by the l.h.s. of Eq.(\ref{stationary3}):
\be
{\cal M}_{v} = \left\{x^\mu\Big|x^\mu = \left({1 \over 2}v^2\Lambda_0-\zeta, \left({1 \over 2}v^2\Lambda-\zeta\right)\vec n\right)\right\}\,,
\label{Malpha}
\ee
with $v^2$ fixed and all possible $\zeta \in \mathbb{R}$ and $\vec n$. This manifold is a surface of a pair of oppositely
oriented 4D cones, whose common vertex is located at ${1 \over 2}v^2(\Lambda_0 - \Lambda, \vec 0)$ lying
on the future line.

As $v^2$ grows, ${\cal M}_{v}$ is homothetically rescaled. The shape of the cones remains the same,
their common vertex sliding upwards. When $v^2$ changes from zero to infinity, each of the cones
sweeps a region in the 4D space. The upper cones sweeps the interior of $LC^+$, while the lower cone sweeps
all the space outside $LC^-$. Therefore, if a point is chosen inside $LC^+$, it will be swept twice: once by each cone;
a point outside $LC^+$ and $LC^-$ will be swept only once, by the lower cone. This gives the number of solutions
in the toy model.

A slightly different way to arrive at this conclusion is to consider an ``equal-time" 3D section of ${\cal M}_{v}$
that passes through the point $M^\mu$,
\be
\mu_{v} = \left\{x^\mu\Big|x^\mu = \left(M_0,\, \left[M_0 - {1 \over 2}v^2(\Lambda_0-\Lambda)\right]\vec n\right)\right\}\,,
\label{Malpha2}
\ee
and see how it changes when $v^2$ grows from zero to infinity.
If $M_0$ is fixed and positive, $\mu_{v}$ is a sphere
whose radius shrinks from $M_0$ to zero and then swells all the way up to infinity.
Points that lie inside the initial sphere (which correspond to $M^\mu$ lying inside $LC^+$)
will be swept twice; points outside that sphere will be swept only once. If $M_0$ is negative, then
the above section is a sphere whose radius starts from $|M_0|$ and increases to infinity.
A point outside the initial sphere will be swept once, while a point inside it will
never be crossed.

\subsection{Geometric constructions in the general case}

We now turn to the general case with all $\Lambda_i$ different. Without lack of generality, suppose that
$\Lambda_3 > \Lambda_2 > \Lambda_1$.

Again, consider the sphere $r^\mu = {1 \over 2}v^2(1,\,\vec n) = {1 \over 2}v^2(1,\,n_i)$ for fixed $v^2$. Upon action of
$\Lambda^{\mu\nu}$ it is mapped to the (surface of) ellipsoid ${1 \over 2}v^2(\Lambda_0,\,\Lambda_i n_i)$.
The 3D manifold
\be
{\cal M}_{v} = \left\{x^\mu\Big|x^\mu = \left({1 \over 2}v^2\Lambda_0-\zeta,\,
\left({1 \over 2}v^2\Lambda_1 - \zeta\right) n_1,\,
\left({1 \over 2}v^2\Lambda_2 - \zeta\right) n_2,\,
\left({1 \over 2}v^2\Lambda_3 - \zeta\right) n_3\right) \right\} \label{Malpha3}
\ee
has a more involved geometry. It is composed of 4 parts, which are ``glued" along
2D ``equal time" regions: the upper cone-like manifold, ${\cal M}_v^{(\vee)}$,
which corresponds to $\zeta \in (-\infty,{1 \over 2}v^2\Lambda_1]$,
two compact cusped manifolds, ${\cal M}_v^{(1)}$ and ${\cal M}_v^{(2)}$,
which correspond to $\zeta \in [{1 \over 2}v^2\Lambda_1,{1 \over 2}v^2\Lambda_2]$
and $\zeta \in [{1 \over 2}v^2\Lambda_2,{1 \over 2}v^2\Lambda_3]$, respectively,
and the lower cone-like manifold, ${\cal M}_v^{(\wedge)}$
corresponding to $\zeta \in [{1 \over 2}v^2\Lambda_3,+\infty)$.

\begin{figure}[!htb]
   \centering
\includegraphics[width=7cm]{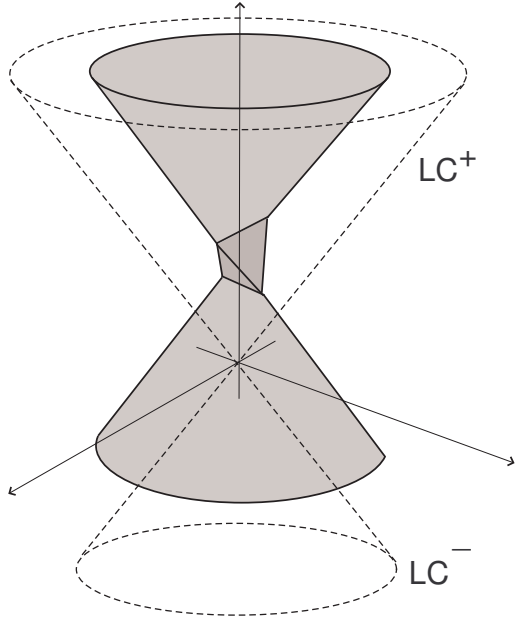}
\caption{A schematic illustration of the lower-dimensional analogue of ${\cal M}_{v}$, (\ref{Malpha3}), for some $v^2 >0$.
In this case it is a 2D ruled surface embedded into 3D space; it is made of the upper cone,
the lower cone and one compact manifold that links them. The $M_0 = \mathrm{const}$ sections are ellipses.}
   \label{fig-cones}
\end{figure}

To help visualize this construction,
we schematically show in Fig.~\ref{fig-cones} a similar warped surface in three-dimensional space, where
there is only one compact manifold that links the two cones.

The 2D regions along which these parts of ${\cal M}_{v}$ are glued, are ellipses together with their interiors.
For example, the border between ${\cal M}_v^{(\vee)}$ and ${\cal M}_v^{(1)}$ is an ellipse
with semiaxes $v^2(\Lambda_2-\Lambda_1)$ and $v^2(\Lambda_3-\Lambda_1)$ that lies in the plane orthogonal to the first axis.

Now consider the family of all ${\cal M}_{v}$ for $v^2 = [0,\infty)$.
As $v^2$ changes, the above construction is rescaled.
As before, the upper cone sweeps the interior of $LC^+$ and the lower cone sweeps all the region outside $LC^-$.
In addition, each of the two compact manifolds sweeps a certain cone in the Minkowski space.
These cones are the {\em caustic 3D surfaces} of the family of half-planes given by the l.h.s. of Eq.~(\ref{stationary3}).
In order to visualize these caustics cones and develop intuition, we suggest to take a look
at a simple planimetric problem described in Appendix~\ref{appsection-caustic}.
Returning then to 2HDM, consider again $\mu_{v}$, the 3D section of ${\cal M}_{v}$ that passes through $M^\mu$.
It is an ellipsoid with semiaxes
\be
M_0 - {1 \over 2}v^2(\Lambda_0-\Lambda_1),
\quad M_0 - {1 \over 2}v^2(\Lambda_0-\Lambda_2),
\quad M_0 - {1 \over 2}v^2(\Lambda_0-\Lambda_3)\,.
\label{semiaxes}
\ee
If $M_0$ is negative, the situation remains the same as before. As $v^2$ changes from zero to infinity,
the ellipsoid defined by semiaxes (\ref{semiaxes}) starts from sphere with radius $|M_0|$ and grows to infinity,
sweeping exactly once each of the points that lie outside the initial sphere.
Thus, if $M^\mu$ in the prototypical model lies inside the past lightcone $LC^-$, there are
no non-trivial stationary points of the potential, in agreement with Proposition~\ref{prop-exist}.

For $M_0>0$ there is a richer set of possibilities. Consider the three special values of $v^2$,
$$
{1\over 2}v^2_i = {M_0 \over \Lambda_0 - \Lambda_i}\,,\quad i=1,2,3\,, \quad 0< v^2_1<v^2_2<v^2_3\,,
$$
when $i$-th semiaxis of $\mu_{v}$ becomes zero.
As long as $0< v^2 < v^2_1$, the 3D section passes through ${\cal M}_v^{(\vee)}$.
It is an the ellipsoid that starts from the initial sphere, gradually shrinks, and at
$v^2 = v_1^2$ it collapses to an ellipse with semiaxes
$$
M_0{\Lambda_2 - \Lambda_1 \over \Lambda_0 - \Lambda_1}\,,\quad
M_0{\Lambda_3 - \Lambda_1 \over \Lambda_0 - \Lambda_1}\,,
$$
that lies in a plane orthogonal to the first axis. Then, for
$v^2_1 < v^2 < v^2_2$ the section passes through ${\cal M}_v^{(1)}$;
the first semiaxis of $\mu_v$ grows while the other two shrink, and at
$v^2 = v_2^2$ it collapses to another ellipse orthogonal to the second axis.
Later, at $v^2 = v_2^2$, $\mu_v$ collapses to the third ellipse,
and after that it keeps swelling up to infinity.

\begin{figure}[!htb]
   \centering
\includegraphics[width=7cm]{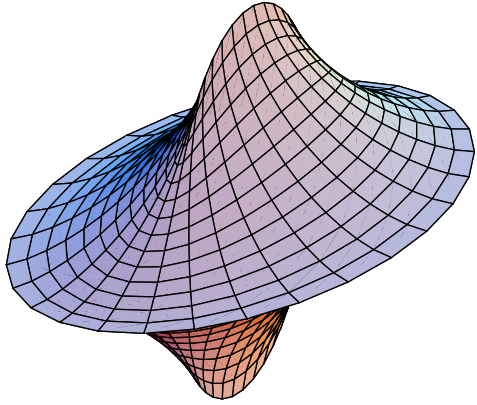}
\caption{The $x_0=\mathrm{const}$ section of one of the cones of caustics. It is a 3D surface that envelops the family of ellipsoids
(\ref{semiaxes}) for $v^2_1 < v^2 < v^2_2$.}
   \label{fig-astroidlike}
\end{figure}

The 3D section of the first caustic cone generated by ${\cal M}_v^{(1)}$ is the region in $\mathbb{R}^3$ swept by
a family of ellipsoids with semiaxes (\ref{semiaxes}) for $v^2_1 < v^2 < v^2_2$. This
region has a cusped astroid-like shape\footnote{In 2D case it would be simply a rescaled astroid, see Appendix~\ref{appsection-caustic}.},
somewhat elongated along the third axis, shown in Fig.~\ref{fig-astroidlike}.
The section of the other caustic cone has a similar shape, although oriented differently.

One can show that as $v^2$ grows from $v^2_1$ to $v^2_2$, each point inside the figure shown in Fig.~\ref{fig-astroidlike}
is swept exactly twice. The same takes place for points inside the 3D section for the second caustic cone, when
$v^2$ grows from $v^2_2$ to $v^2_3$. In addition, each point in the 3D section considered will be always swept
once for some $v^2 > v^2_3$ and, if $M^\mu$ lies inside $LC^+$, will be swept once more for some $0< v^2 < v^2_1$.

The above construction gives a geometric proof of the following criteria:
\begin{proposition}\label{prop-number}
(1) if $M^\mu$ lies outside $LC^-$, at least one neutral stationary point in the Higgs field space exists;\\
(2) if $M^\mu$ lies inside $LC^+$, at least two neutral stationary points exist;\\
(3) if $M^\mu$, in addition, lies inside one of the caustic cones, two additional neutral stationary points appear;\\
(4) if $M^\mu$ lies inside both caustic cones, four neutral stationary points appear, in addition to criteria (1) or (2).
\end{proposition}
Thus, the overall number of nontrivial stationary points can be as high as six neutral plus one charge-breaking,
in agreement with results of \cite{maniatis}.
The sequence of these solutions, which are given by intersections of the evolving $\mu_v$ with a given point $M^\mu$,
is easy to visualize. For example, consider a generic $M^\mu$ that lies sufficiently close to the future-directed
eigenvector of $\Lambda^{\mu\nu}$. Then, it lies inside $LC^+$ and both caustic cones.
As the ellipsoid starts from sphere and shrinks, it will pass this point once before collapsing to the first ellipse.
Then, evolving between the first and the second mutually orthogonal ellipses, it will pass the point twice.
Then, evolving from the second to the third ellipses it will yield two more intersections, and
finally, swelling from the last ellipse to infinity it will pass the point the sixth time.

A special discussion is needed when $M^\mu$ is not a generic point,
but lies inside one of the ellipses described above, say, the first one.
The planimetric problem described
in Appendix~\ref{appsection-caustic} helps understand what happens in this case.
Instead of two intersections for values of $v^2$ just below and just above $v_1^2$,
there are two different solutions at $v^2 = v_1^2$,
so that the counting rules stated in Proposition~\ref{prop-number} remain unchanged.
These two solutions correspond to non-zero and opposite values of the coordinate $r_1$, breaking thus the
$\vec e_1 \to -\vec e_1$ reflection symmetry of $\Lambda^{\mu\nu}$ and $M^\mu$.

Note that such a breaking of discrete symmetry occurs if and only if the vector $M^\mu$ lies inside
at least one of the caustic cones.
The spontaneous $CP$-violation in 2HDM to be discussed in the next section has
precisely these geometric roots.

Note that we analyzed above the most general case. Partially degenerate situations
(e.g. $\Lambda_1 = \Lambda_2$, etc.) can be analyzed in a similar way.

Proposition~\ref{prop-number} gives the number of neutral stationary points of a given Higgs potential,
but does not say anything about their minimum/saddle point nature. As was showed before,
to establish that a given stationary point is a minimum, one needs to find all the eigenvalues of the mass matrix
(\ref{massNeutral4}) and make sure that all of them are positive. Whether this
criterion admits a simple geometric formulation, remains to be seen.

\section{Spontaneous $CP$-violation}\label{sectionCP}

The Higgs potential is said to be explicitly $CP$ conserving, if it commutes
with the operator of the $CP$ transformation. It is known that the Higgs potential in 2HDM
is explicitly $CP$-conserving if and only if there exists such a $SU(2)$ reparametrization transformation that
makes all the coefficients real, see discussion in \cite{haber2}. However, in this case it is
possible that the vacuum expectation values for the two doublets attain a non-zero relative phase,
which, in general, cannot be ``rotated away" by a unitary transformation
of the fields still keeping the potential real (for two examples
of models when this phase {\em can} be rotated away, see \cite{Branco,LLY}).
When this happens, the $CP$-conservation is broken spontaneously.
In fact, this possibility was the original motivation to consider EWSB with two Higgs doublets, \cite{2HDM}.

Undoubtedly, for any explicitly $CP$-conserving Higgs potential with a given set of $\lambda_i$ and
$m_{ij}^2$ there exists a unique answer to the question whether spontaneous $CP$ violation occurs.
However, up to now no such criterion for a general 2HDM has been published.
What one encounters in literature is either straightforward minimization of the Higgs potential in some
simple cases, \cite{2HDM,Branco,LLY}, or discussion of what would be the properties of the $CP$-conserving or $CP$-violating
vacuum solutions, if they exist, \cite{haber2,nishi}. For example, \cite{haber2} contains
a set of basis-invariant conditions that must be satisfied in order to guarantee that no spontaneous
$CP$-violation occurs. Aside from very complicated algebraic expressions for these conditions,
they are presented in implicit form: in order to check if a given 2HDM exhibits spontaneous $CP$ violation,
one would need first to {\em explicitly} find the minima and then check if they satisfy the algebraic criteria.

Here we tackle this problem with the geometric approach developed above.

\subsection{Preliminary conditions for spontaneous $CP$-violation}

Let us return for a moment to the conventional $SU(2)$-reparametrization group, which
generates 3D rotations in the orbit space.
Here, instead of $\Lambda_{\mu\nu}$ one deals with scalar $\Lambda_{00}$, vector $\Lambda_{0i}$ and rank-two symmetric tensor
$\Lambda_{ij}$, \cite{ivanov,nishi}. The four-vector $M_\mu$ is split into scalar $M_0$ and
3D vector $M_i$.

As shown in \cite{ivanov}, the basis-invariant condition for the existence of a ``hidden" real potential form
(i.e. for the explicit $CP$-conservation), is that among the three eigenvectors of $\Lambda_{ij}$
there must be one orthogonal both to $\Lambda_{0i}$ and $M_i$. We now extend this statement to the spontaneous $CP$-violation:
\begin{proposition}\label{prop-noCP}
Spontaneous $CP$-violation can take place only if there exists one and only one
eigenvector of $\Lambda_{ij}$ orthogonal both to $\Lambda_{0i}$ and $M_i$.
\end{proposition}

\begin{proof}
The first part is proved in \cite{ivanov}. We now show that if there are {\em two} eigenvectors
of $\Lambda_{ij}$ orthogonal both to $\Lambda_{0i}$ and $M_i$, then spontaneous $CP$-violation cannot take place.

The eigenvectors of a symmetric real matrix $\Lambda_{ij}$ are mutually orthogonal.
If non-zero vectors $\Lambda_{0i}$ and $M_i$ happen to be orthogonal to two eigenvectors
of $\Lambda_{ij}$, then they must be proportional to the {\em third} eigenvector of
$\Lambda_{ij}$. In order words, $\Lambda_{0i}$ and $M_i$ are {\em themselves}
eigenvectors of $\Lambda_{ij}$. As shown in \cite{ivanov}, this situation corresponds precisely
to the existence of hidden $Z_2$-symmetry in the Higgs potential.
Diagonalizing $\Lambda_{ij}$, one arrives at
$$
\Lambda_{ij} =
\left(\begin{array}{ccc}
\Lambda_{11} & 0 & 0 \\
0 & \Lambda_{22} & 0 \\
0 & 0 & \Lambda_{33} \end{array}\right)\,,\quad
\Lambda_{0i} = (0,\,0,\,\Lambda_{03})\,,\quad
M_i = (0,\, 0,\, M_3)\,.
$$
We are now going to prove the following statement:
even if one manages to find a solution with $r_2 \equiv 2\Im(\phi_1^\dagger \phi_2) \not = 0$,
then one can always ``rotate away" the relative phase still keeping the parameters real.
Indeed, let us write the first and second components of the equation $\Lambda_{\mu\nu} n^\nu v^2/2 - \zeta n_\mu = M_\mu$:
$$
\left({v^2 \over 2}\Lambda_{11} -\zeta\right) n_1 = 0\,,\quad \left({v^2 \over 2}\Lambda_{22} -\zeta\right) n_2 = 0\,.
$$
Two possibilities are to be considered: $\Lambda_{11} = \Lambda_{22}$ and $\Lambda_{11} \not = \Lambda_{22}$.
The first possibility implies also a possibility to have both $n_1$ and $n_2$ simultaneously non-zero. However,
the matrix $\Lambda_{ij}$ possesses in this case an extra symmetry that allows one to manually align the first axis along the
transverse vector $\vec n_\perp$. This rotation sets $n_2 = 0$ and removes the relative phase between the two v.e.v.'s.
For the second possibility, one can have only one of $n_1$ and $n_2$ nonzero, but not both. If this non-zero
component happens to be $n_2$, then $n_1$ is necessarily zero. One can then flip the first and second axes
and arrive at the real Higgs potential with $n_2 = 0$.
We conclude that there is no room for spontaneous $CP$-violation in this case.
\end{proof}

Put in simple terms, this Proposition means that in order to exhibit not explicit, but spontaneous $CP$-violation,
the potential must have certain symmetry, but it must not be {\em too symmetric}.
It might be interesting to see if this is a generic rule that is observed when a spontaneous breaking
of a (discrete) symmetry takes place.

The fact that the non-zero phase does not always imply spontaneous $CP$-violation has been known since long
ago (see discussion in \cite{haber2}).
For example, authors of \cite{Branco} considered a 2HDM with $\lambda_1 = \lambda_2$, $\lambda_6 = \lambda_7 = $ real, $\lambda_5$ real,
$m_{11}^2 = m_{22}^2$, and $m_{12}^2$ real, and showed that the relative phase between the v.e.v.'s can be rotated away
still keeping all the parameters real. In our language, this model corresponds to $M_i = (M_1,\, 0,\, 0)$, $\Lambda_{0i} = (\Lambda_{01},\,
0,\, 0)$ and diagonal $\Lambda_{ij}$, which obviously violates conditions of Proposition~\ref{prop-noCP}.
Paper \cite{LLY} deals with an even simpler example: $m_{12}^2 = \lambda_6 = \lambda_7 =0$, which literally corresponds
to the situation just discussed in the proof of this Proposition.
Our linear algebraic approach reveals the origin of these sporadic findings.

\subsection{Necessary and sufficient conditions for spontaneous $CP$-violation}

Given that the requirements stated in Proposition~\ref{prop-noCP} are satisfied,
the following questions now arise: (1) what are the necessary and sufficient conditions for existence
of a spontaneous $CP$-violating stationary points and (2) when this stationary point is a minimum?

The answer to the first question is given by the following Proposition:
\begin{proposition}\label{prop-CP}
Given that the conditions of Proposition~\ref{prop-noCP} are satisfied,
the spontaneously $CP$-violating stationary points exist if and only if
the components $M^\mu$ of the prototypical model satisfy the following inequality:
\be
{M_1^2 \over (\Lambda_1-\Lambda_2)^2} + {M_3^2 \over (\Lambda_3-\Lambda_2)^2} < {M_0^2 \over (\Lambda_0-\Lambda_2)^2}\,.
\label{CPconditions}
\ee
\end{proposition}

\begin{proof}
As already shown at the end of the previous section, in order for the {\em prototypical model} to exhibit
spontaneous $CP$-violation, $M^\mu$ must lie inside an appropriate caustic cone and on the plane orthogonal to the second axis.
Such a non-trivial caustic cone exists if and only if $\Lambda_2$ is a non-degenerate eigenvalue of $\Lambda^{\mu\nu}$.
If this is satisfied, $M^\mu$ must lie inside the ellipse with semiaxes
$M_0|\Lambda_1-\Lambda_2|/(\Lambda_0-\Lambda_2)$ and $M_0|\Lambda_3-\Lambda_2|/(\Lambda_0-\Lambda_2)$,
which gives (\ref{CPconditions}).

To complete the proof, one needs to show that the passage from the original to the prototypical model and back respects
the $CP$-symmetry properties. We start with the Higgs potential that has (hidden) explicit $CP$-conservation
and bring it to the real potential form
by an $SU(2)$ reparametrization transformation. The corresponding $SO(3)$ rotation in the orbit space
brings the tensor $\Lambda^{\mu\nu}$ and 4-vector $M^\mu$ to the following form:
$$
\Lambda^{\mu\nu} =
\left(\begin{array}{cccc}
\cdot & \cdot & 0 & \cdot\\
\cdot & \cdot & 0 & \cdot\\
0 & 0 & -\Lambda_2 & 0\\
\cdot & \cdot & 0 & \cdot \end{array}\right)\,,\quad
M^\mu = (\cdot,\ \cdot,\ 0, \ \cdot)\,,
$$
where dots represent generic values. Diagonalization of $\Lambda^{\mu\nu}$ requires now
only boosting along the first and third axes and 2D rotation between them.
Such transformations do not generate the second components of $\Lambda_{0i}$ or $M_i$,
therefore the prototypical model is also explicitly $CP$-conserving.

The inverse transformation laws, too, do not involve the second axis.
If one obtains a (non)zero value of $r_2$ in the prototypical model, this value
remains the same upon returning to the original Higgs potential. Within the conditions of
Proposition~\ref{prop-noCP}, the non-zero phase cannot be ``rotated away",
and the spontaneous $CP$-violation takes place in the original model if and only if
it takes place in the prototypical model.
\end{proof}

In Appendix~\ref{appsection-direct} we rederive condition (\ref{CPconditions}) with the aid of straightforward algebra.
Unfortunately, it appears to be a non-trivial task to check if the conditions derived in \cite{haber2}
for the spontaneous $CP$-violation in a generic explicitly $CP$-conserving model agree with ours.

Appendix~\ref{appsection-direct} lists also the necessary and sufficient conditions for the
spontaneously $CP$-violating point to be a minimum. Here we cite only the particularly simple
necessary condition: {\em spontaneously $CP$-violating stationary point of the potential can be a minimum only if
$\Lambda_2$ is the largest spacelike eigenvalue of $\Lambda^{\mu\nu}$}.

When a $CP$-violating minimum is the global one still remains to be studied.

\section{Discussion and conclusions}

In this work we exploited the simple observation that the generic Higgs potential of the two-Higgs-doublet model
keeps its form under any linear transformation between the two doublets.
The most interesting subgroup $SL(2,C)$ of this group of transformations induces the Lorentz group of transformations
of $r^\mu = \Phi^\dagger \sigma^\mu \Phi$. The Higgs potential takes a very compact and Lorentz invariant form
$V = - M_\mu r^\mu + {1\over 2}\Lambda_{\mu\nu} r^\mu r^\nu$.
This extended reparametrization symmetry turns out very useful in analysis
of the existence and number of extrema of the Higgs potential and their classification according to
neutral/charge-breaking and saddle/minimum nature.

We introduced the notion of a {\em prototypical model}, which has only four free parameters in the quartic potential.
All 2HDM's linked to the prototypical model by an $SL(2,C)$ transformation, share the same
set of solutions to the stationary point problem. This construction gives the answer to
the question which free parameters are physically essential and which are redundant when we
study general properties of 2HDM.

It has always seemed that explicit manipulation with high-order algebraic equations are unavoidable
in order to learn something about the vacuum in 2HDM.
The key point of the present work is the observation that we {\em do not need}
such manipulations to understand the general structure of the 2HDM vacuum.

Several geometric constructions appear naturally in the orbit space of 2HDM:
the future lightcone $LC^+$, its image under the action of $\Lambda^{\mu\nu}$, and the caustic cones.
Using them, we proved that the charge-breaking and neutral minima cannot coexist in 2HDM.
We also discovered the role of the caustic manifolds as separatrices between the 2HDM's with different number of solutions.
Finally, we established the geometric roots of the spontaneous breaking of discrete symmetries
and obtained explicit conditions when spontaneous $CP$-violation happens in 2HDM.
What remains to be seen is when these spontaneously violating minima are the global ones.

A number of directions for further study opens up.
\begin{itemize}
\item
In the present work we did not consider in detail different types of degenerate situations.
It seems interesting to classify such situations in terms of little groups of residual symmetries,
and trace down the consequences for the 2HDM vacuum. This study can be done within the geometric framework
developed here, without any need of explicit expressions for positions of extrema.
\item
If $\lambda_i$ are fixed and $m_{ij}^2$ continuously change,
the global minimum undergoes certain zero-temperature phase transitions. Using geometric approach,
one can classify the phase transitions allowed in 2HDM and study their properties.
\item
Check if the geometric approach can be applied also to Higgs-Higgs scattering processes.
In particular, search for a simpler rederivation of the perturbative unitarity constraints,
 \cite{GIunitarity}.
\item
Study the deformation of the above description when quantum corrections to the potential
are taken into account.
\item
Extend the geometric approach to other forms of the Higgs
potential, for example, $N$-Higgs doublet model. Indeed, in NHDM
one can again construct a scalar and an $N^2-1$-component vector
in the adjoint representation of $SU(N)$, see \cite{nishi}. Then,
switching from reparametrization group $SU(N)$ to $SL(N,C)$, one
can unite them in a single vector in $1+(N^2-1)$-space with
Minkowski structure. This should provide extra aid to attempts to
understand the general properties of NHDM, such as
\cite{barroso2006}. Finally, the formalism developed here might
turn out useful for analysis of Landau-Ginzburg potentials in some
condensed matter problems.
\end{itemize}

I am thankful to Ilya Ginzburg, Roman Li, Celso Nishi, Alessandro Papa, Gianfranco Sartori, Gianpaolo Valente
and the anonymous referee for discussions and useful comments.
This work was supported by the INFN and FNRS,
and partly by grants RFBR 05-02-16211 and NSh-5362.2006.2.
\appendix

\section{Diagonalization of $\Lambda^{\mu\nu}$}\label{appsection-lambda}

We first show some basic facts on diagonalization of the real symmetric tensor $\Lambda^{\mu\nu}$ in the Minkowski space.
$\Lambda^{\mu\nu}$ can be viewed as an operator acting on vectors in the Minkowski space.
The (right) eigenvector $p^\mu$ with eigenvalue $\Lambda$ is defined by
$$
\Lambda^{\mu\rho} g_{\rho\nu} p^\nu = \Lambda p^\mu\,,
$$
thus the characteristic equation for the eigenvalues is $\det(\Lambda^{\mu\nu} - \Lambda g^{\mu\nu}) = 0$.
One can switch to the more common expression by lowering one of the superscripts:
\be
\det(\Lambda^{\mu}{}_\nu - \Lambda \delta^{\mu}_{\nu}) = 0\,. \label{char}
\ee
Tensor $\Lambda^{\mu}{}_\rho$ that enters (\ref{char}) is still real, but {\em not symmetric} anymore.
Therefore, its eigenvalues are, in general, complex. If this is the case, the transformation that diagonalizes
$\Lambda^{\mu\nu}$ would involve a complex transformation matrix, which does not belong
to the $SO(1,3)$ group. This means that exploiting only the proper Lorentz group of transformations,
one might not be able to diagonalize a given $\Lambda^{\mu\nu}$.

The situation simplifies due to the requirement that $\Lambda^{\mu\nu}$ is positive definite on the future
lightcone\footnote{I am thankful to Celso Nishi who discovered a flaw in the earlier version of the proof
of this proposition}:

\begin{proposition}\label{appA}
Tensor $\Lambda^{\mu\nu}$ is positive definite on the future lightcone if and only if the following three conditions
hold:\\
(1) $\Lambda^{\mu\nu}$ is diagonalizable by an $SO(1,3)$ transformation,\\
(2) the timelike eigenvalue $\Lambda_0$ is positive,\\
(3) all spacelike eigenvalues $\Lambda_i$ are smaller than $\Lambda_0$.
\end{proposition}

\begin{proof}
Obviously, if $\Lambda^{\mu\nu}$ satisfies conditions (1)--(3), then the positive definiteness follows immediately.
So, one needs to prove that conditions (1)--(3) do follow from the positive definiteness.

Despite $\Lambda_{\mu\nu}$ being real and symmetric, its
eigenvalues can be complex because of the non-Euclidean metric.
The first step is to prove that the positive definiteness in the
future lightcone $LC^+$ implies that all the eigenvalues of
$\Lambda_{\mu\nu}$ are real.

Indeed, suppose there is a pair of complex eigenvalues, $\lambda$ and $\lambda^*$,
with respective complex eigenvectors $p^\mu$ and $q^\mu$:
$$
{\Lambda^{\mu}}_{\nu} p^\nu = \lambda p^\mu\,,\quad
{\Lambda^{\mu}}_{\nu} q^\nu = \lambda^* q^\mu\,.
$$
One can show that there can be only one pair of complex eigenvalues, thus, $\lambda$
is non-degenerate.
Since $\Lambda^{\mu}_{\nu}$ is real, $q^\mu \propto p^{\mu *}$ (and can be taken equal to $p^{\mu *}$).
These eigenvectors are orthogonal, $p^\mu q_\mu = 0$, (it follows from the standard
argument due to $\lambda \not = \lambda^*$) and can be normalized so that
$p^\mu p_\mu = q^\mu q_\mu = 1$.

Consider now a real vector $r^\mu$
$$
r^\mu = c p^\mu + c^* p^{* \mu}\,.
$$
Suppose that
$r^\mu r_\mu = c^2 + c^{* 2} = 2|c|^2\cos(2\phi_c) > 0$,
so that either $r^\mu$ or $-r^\mu$ lies inside
the forward lightcone. Then, the corresponding quadratic form is
$$
\Lambda_{\mu\nu} r^\mu r^\nu = \lambda c^2 + \lambda^* c^{* 2}
\equiv 2|\lambda||c|^2\cos\left(2\phi_c + \phi_\lambda\right)\,.
$$
Due to the phase shift $\phi_\lambda \not = 0$ one can always find $\phi_c$
such that $\cos(2\phi_c) > 0$ but $\cos(2\phi_c + \phi_\lambda)<0$,
i.e. one can always find an $r^\mu \in LC^+$ such that $\Lambda_{\mu\nu} r^\mu r^\nu < 0$,
which contradicts the assumption.

Since all the eigenvalues of $\Lambda_{\mu\nu}$ are real, the eigenvectors also
can be chosen all real and orthogonal.
One can show that they can be normalized so that one of the eigenvectors has positive norm,
$p_0^\mu p_{0 \mu} = 1$, while the other three have negative norms $p_i^\mu p_{i \mu} = -1$
for each $i=1,2,3$.
Thus, the transformation matrix $T$ that diagonalizes $\Lambda_{\mu\nu}$
is real, and after diagonalization $\Lambda_{\mu\nu}$ takes form
$\mathrm{diag}(\Lambda_0,\, -\Lambda_1,\, -\Lambda_2,\, -\Lambda_3)$.
Note that transformation $T$ also conserves norm, $r^\mu r_\mu = const$.
It means that $T$ can be realized as a transformation from the proper Lorentz group.

Now, the requirement that $\Lambda^{\mu\nu}$ is positive definite in $LC^+$ means
$$
\Lambda_0 - \rho(\Lambda_1 \sin\theta\cos\phi + \Lambda_2 \sin\theta\sin\phi + \Lambda_3\cos\theta) > 0
$$
for all $0<\rho<1$, $0\le \theta\le \pi$ and all $\phi$. This holds when $\Lambda_0$ is positive and is larger than all $\Lambda_i$.
\end{proof}

Note that since $\Lambda^{\mu\nu}$ is a linear operator, it maps points inside some surface $S$ into points
that also lie inside the image of this surface $S^\prime$.

The positivity conditions in the form of inequalities among $\lambda$'s can be readily
obtained from this statement. Here we reproduce the standard set of positivity conditions for
a model with explicit $Z_2$-symmetry, i.e. when $\lambda_6 = \lambda_7 = 0$.
$\Lambda_{\mu\nu}$ of this model has form
$$
\Lambda^{\mu\nu} = {1\over 2}\left(\begin{array}{cccc}
{\lambda_1+\lambda_2 \over 2} + \lambda_3 & 0 & 0 & -{\lambda_1-\lambda_2 \over 2} \\[1mm]
0 & \lambda_4 + \Re\lambda_5 & -\Im\lambda_5 & 0 \\[1mm]
0 & -\Im\lambda_5 & \lambda_4-\Re\lambda_5 & 0 \\[1mm]
 -{\lambda_1-\lambda_2 \over 2} & 0 & 0 & {\lambda_1+\lambda_2 \over 2} - \lambda_3
\end{array}\right)\,,
$$
which after a rotation among the first and second axes and an appropriate boost along the third axis gives
$$
\Lambda^{\mu\nu} = {1\over 2}\left(\begin{array}{cccc}
\lambda_3 + \sqrt{\lambda_1\lambda_2} & 0 & 0 & 0 \\[1mm]
0 & \lambda_4 + \lambda_5 & 0 & 0 \\[1mm]
0 & 0 & \lambda_4-\lambda_5 & 0 \\[1mm]
 0 & 0 & 0 & \sqrt{\lambda_1\lambda_2} - \lambda_3
\end{array}\right)\,.
$$
Such a transformation is possible if and only if $\lambda_1+\lambda_2 > |\lambda_1-\lambda_2|$,
which is equivalent to $\lambda_1>0$, $\lambda_2>0$. The eigenvalues of $\Lambda_{\mu\nu}$ are
$$
2 \Lambda_0 = \lambda_3 + \sqrt{\lambda_1\lambda_2}\,,\quad
2 \Lambda_1 = - \lambda_4 - |\lambda_5|\,,\quad
2 \Lambda_2 = - \lambda_4 + |\lambda_5|\,,\quad
2 \Lambda_3 = \lambda_3 - \sqrt{\lambda_1\lambda_2}\,.
$$
The conditions stated in Proposition~\ref{appA} now translate into
$$
\lambda_1>0\,,\quad \lambda_2>0\,,\quad
\lambda_3 + \sqrt{\lambda_1\lambda_2} > 0\,,\quad
\lambda_3 + \lambda_4 - |\lambda_5| + \sqrt{\lambda_1\lambda_2} > 0\,,
$$
which coincides with the standard set of inequalities.

\section{Planimetric primer: caustics, number of solutions and spontaneous violation of discrete symmetries}\label{appsection-caustic}
The following simple planimetric problem provides insight into phenomena that take
place in the original problem.

Given a linear map of the 2D vector space
$r_i \to r^\prime_i = A_{ij} r_j - B_i$ with real symmetric matrix $A_{ij}$ and real
$B_i$, how many points are there on the unit circle such that $r^\prime$ is proportional to $r$?

\begin{figure}[!htb]
   \centering
\includegraphics[height=7cm]{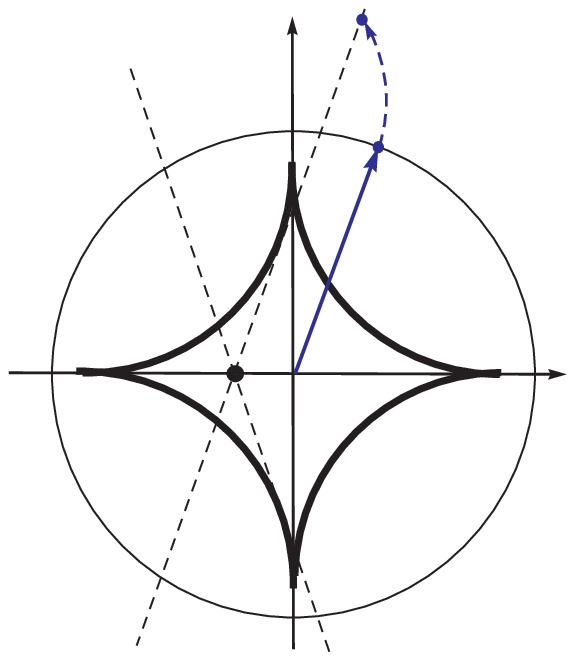}
\includegraphics[height=7cm]{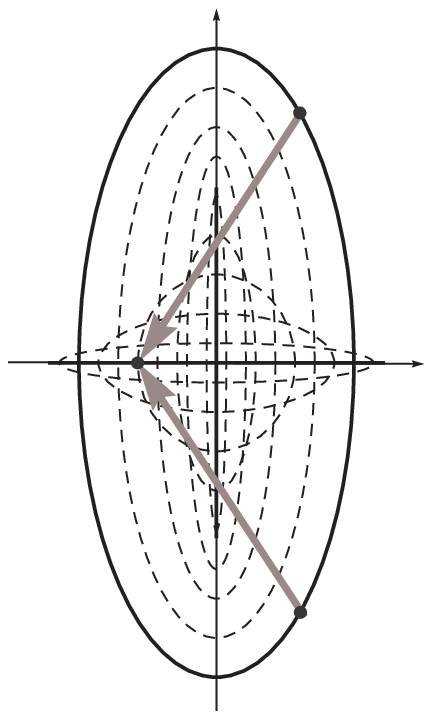}
\caption{(Left) The caustic curve formed by the family of straight lines $A_{ij}r_j - \zeta \cdot r_i$ for $r_i$ lying on the unit circle.
A point in the region bounded by this curve has four lines passing through it. For a point $B_i$ deliberately chosen
on the horizontal axis, there are nontrivial solutions, one of which is also shown, that do not lie on it. (Right)
The same solution visualized as a shrinking ellipse. When an ellipse collapse onto the line, on which $B_i$ lies,
two points from $A_{ij}r_j$ arrive at $B_i$.}
   \label{fig-caustics}
\end{figure}

Following the strategy outlined in Section~\ref{section-strategy},
take a point $r_i$ on the unit circle and draw a straight line $A_{ij}r_j - \zeta \cdot r_i$ trying to arrive at $B_i$.
Repeating this for every point on the unit circle, one gets a family of straight lines.
If $A_{ij}$ is not proportional to the unit matrix, these lines do not intersect at a single point,
but instead form a {\em caustic curve} (an astroid in the present case). If $B_i$ is inside the region bounded
by the caustic curve, there are four solutions; if $B_i$ is outside the caustics curve, there are only two solutions.
Fig.~\ref{fig-caustics},~left, illustrates this construction. For a point inside the astroid deliberately chosen
to lie on the horizontal axis, there are two ``trivial" solutions, which also lie on this axis, and two nontrivial solutions,
one of which is also shown.

Although both $A_{ij}$ and $B_i$ are symmetric under flip of the vertical axis,
the two non-trivial solutions are not. Thus, having $B_i$ inside the caustic curve makes spontaneous
breaking of a discrete symmetry possible.

In order to understand better the origin of this phenomenon, consider another way to visualize
the solution of this problem. The matrix $A_{ij}$ acting on the circle $r_i$ maps it to an ellipse with semiaxes
$a_1$ and $a_2$, the eigenvalues of $A_{ij}$. As $\zeta$ increases, the ellipse $A_{ij}r_j - \zeta \cdot r_i$
shrinks, see Fig.~\ref{fig-caustics},~right. At $\zeta = a_1$ and $\zeta = a_2$ this ellipse collapses to an interval
on the vertical and horizontal axes, respectively. In the latter case, in particular, one obtains $B_2 = 0$
without requiring $r_2$ to be zero.

\section{Spontaneous $CP$-violation in the prototypical model: straightforward algebra}\label{appsection-direct}

\begin{figure}[!htb]
   \centering
\includegraphics[height=5cm]{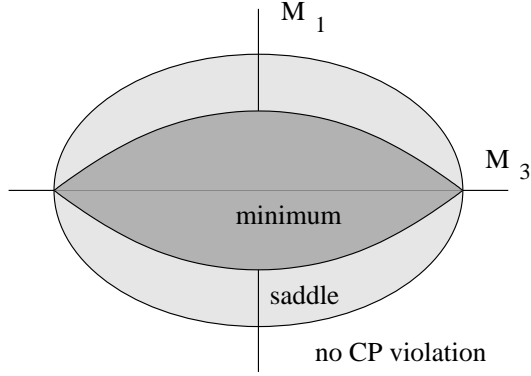}
\caption{Regions on the $M_i = (M_1,\,0,\,M_3)$ plane that lead to spontaneous $CP$-violation
in the particular case $\Lambda_0 = 4$, $\Lambda_1=1$, $\Lambda_2=3$, $\Lambda_3=0$.
If $M_i$ lies outside the outer ellipse, there is no spontaneous $CP$-violation.
If $M_i$ lies inside the darker region, then the $CP$-violating stationary points are minima.
If $M_i$ lies in between, then $CP$-violating stationary points are saddle points.}
   \label{fig-min}
\end{figure}

Here we obtain the criterion (\ref{CPconditions}) using straightforward algebra similar to the
original calculations of \cite{2HDM}. Suppose $\lambda_5 > \lambda_4$, which
guarantees that the non-trivial minima of the potential are neutral and allows us to search the solutions
in form
\be
\phi_1 = {1\over\sqrt{2}}\stolbik{0}{\rho_1}\,,\quad \phi_2 = {1\over\sqrt{2}}\stolbik{0}{\rho_2 e^{i\theta}}\,.
\ee
After substituting them into the Higgs potential of the prototypical model (\ref{Vproto}), we obtain (suppressing barred
notation for simplicity)
\be
V = -{1 \over 4}(m_{11}^2\rho_1^2 + m_{22}^2\rho_2^2 + 2 m_{12}^2\rho_1 \rho_2 \cos\theta)
+ {1\over 8}\left[\lambda (\rho_1^4 + \rho_2^4) + 2 \rho_1^2\rho_2^2(\lambda_3 + \lambda_4 + \lambda_5 \cos 2\theta) \right]\,.
\label{Vsub1}
\ee
This potential can have both $CP$-conserving and $CP$-violating stationary points.
Let us find conditions when spontaneous $CP$-violation takes place.
Since $\sin\theta \not = 0$, we first set $\partial V/\partial \cos\theta =0$, which gives
\be
\cos\theta = {m_{12}^2 \over 2\lambda_5 \rho_1^2\rho_2^2} \equiv \delta\,.
\ee
The condition for non-zero phase is therefore
\be
|\delta| < 1\quad \to \quad {(m_{12}^2)^2 \over \lambda_5^2} < 4 \rho_1^2 \rho_2^2\,.\label{regiondelta}
\ee
The potential can be then rewritten as
\be
V = -{1 \over 4}(m_{11}^2\rho_1^2 + m_{22}^2\rho_2^2)
+ {1\over 8}\left[\lambda (\rho_1^4 + \rho_2^4) + 2 \rho_1^2\rho_2^2\lambda_{345} \right]
+ {1 \over 2}\lambda_5 \rho_1^2\rho_2^2(\cos\theta - \delta)^2 - {(m_{12}^2)^2 \over 8\lambda_5}\,,
\ee
where $\lambda_{345} \equiv \lambda_3 + \lambda_4 - \lambda_5$.
Upon differentiating the potential in respect to $\rho_1^2$ and $\rho_2^2$
(note that both $\rho_1$ and $\rho_2$ are non-zero) and by using $\cos\theta-\delta = 0$,
one obtains:
\be
\rho_1^2 = {m_{11}^2 \lambda - m_{22}^2\lambda_{345} \over \lambda^2 - \lambda_{345}^2}\,,\quad
\rho_2^2 = {m_{22}^2 \lambda - m_{11}^2\lambda_{345} \over \lambda^2 - \lambda_{345}^2}\,.
\ee
Exploiting the identity
$$
4(ax-by)(ay-bx) = (a-b)^2(x+y)^2 - (a+b)^2(x-y)^2\,,
$$
one rewrites (\ref{regiondelta}) as
\be
{(m_{12}^2)^2 \over \lambda_5^2} < {(m_{11}^2+m_{22}^2)^2 \over (\lambda + \lambda_{345})^2} -
{(m_{11}^2-m_{22}^2)^2 \over (\lambda - \lambda_{345})^2}\,.
\ee
This coincides with (\ref{CPconditions}) if one recalls the definition of $M_\mu$ and notes that
$$
\Lambda_0 - \Lambda_2 = {1\over 2}(\lambda+\lambda_{345})\,,\quad \Lambda_2 - \Lambda_3 = {1\over 2}(\lambda-\lambda_{345})\,,\quad
\Lambda_2 - \Lambda_1 = \lambda_5\,.
$$
In order to establish the minimum conditions for the $CP$-violating stationary point, we calculate the matrix of
second derivatives of the potential in respect to $\rho_1^2$, $\rho_2^2$ and $\theta$
\be
D = {1\over 4}\left(\begin{array}{ccc}
\lambda & \lambda_{345} & -\sin\theta\, m_{12}^2{\rho_2 \over \rho_1} \\[2mm]
\lambda_{345} & \lambda & -\sin\theta\, m_{12}^2{\rho_1 \over \rho_2} \\[2mm]
-\sin\theta\, m_{12}^2{\rho_2 \over \rho_1} & -\sin\theta\, m_{12}^2{\rho_1 \over \rho_2} &
4\sin^2\theta\rho_1^2\rho_2^2\lambda_5
\end{array}\right)
\label{secondderiv}
\ee
This matrix is positive definite if and only if
\be
\lambda>0\,,\quad \lambda_5 > 0\,,\quad \lambda^2 - \lambda_{345}^2 > 0\,, \quad\det D > 0\,.\label{lambdascrit}
\ee
From the first three conditions one obtains the following necessary conditions for the spontaneous $CP$-violating point
to be a minimum: $\Lambda_2 > \Lambda_1$ and $\Lambda_2 > \Lambda_3$.
The last condition in (\ref{lambdascrit}) can be rewritten as
\be
{(m_{12}^2)\over \lambda_5^2} < 4 \rho_1^2\rho_2^2 {\lambda^2-\lambda_{345}^2 \over \lambda_5
\left[\lambda\left({\rho_1^2\over\rho_2^2} + {\rho_2^2\over\rho_1^2}\right) - 2\lambda_{345}\right]}\,,\label{regionminimum}
\ee
compare it with (\ref{regiondelta}). This extra condition cuts out a subregion in the ellipse
defined by (\ref{regiondelta}), where the spontaneous $CP$-violating solutions are minima.
Using the parameters of the prototypical model, one can define
$$
x = {M_1 \over M_0} \cdot {\Lambda_0 - \Lambda_2 \over \Lambda_2 - \Lambda_1}\,,\quad
z = {M_3 \over M_0} \cdot {\Lambda_0 - \Lambda_2 \over \Lambda_2 - \Lambda_3}\,.
$$
The condition (\ref{regiondelta}) for the $CP$-violating stationary points is then $x^2+z^2 < 1$, while the condition
(\ref{regionminimum}) for the $CP$-violating minimum takes the form
\be
\left({\Lambda_2 - \Lambda_1 \over \Lambda_0 - \Lambda_2} + {\Lambda_2 - \Lambda_1 \over \Lambda_2 - \Lambda_3}\, z^2 \right)
x^2 < (1-z^2)^2\,.
\ee
Fig.~\ref{fig-min} illustrates the shape of this region in a representative case
$\Lambda_0 = 4$, $\Lambda_1=1$, $\Lambda_2=3$, $\Lambda_3=0$.

Note that checking whether the $CP$-violating minimum is a global one is still a non-trivial task
even in the prototypical model.


\begin{thebibliography}{99}
\bibitem{Hunter}
J.F.~Gunion, H.E.~Haber, G.~Kane, S.~Dawson, {\em The Higgs
Hunter's Guide} (Addison-Wesley, Reading, 1990).
\bibitem{djouadi1}A.~Djouadi, hep-ph/0503172.
\bibitem{2HDM}T.~D.~Lee, Phys.\ Rev.\ D {\bf 8} (1973) 1226.
\bibitem{Sanchez} R.A.~Diaz Sanchez, Ph.~D. Thesis, hep-ph/0212237.
\bibitem{ginzreview}I.~F.~Ginzburg and M.~Krawczyk,  Phys.\ Rev.\ D {\bf 72} (2005) 115013.
\bibitem{haber}S.~Davidson and H.~E.~Haber,  Phys.\ Rev.\ D {\bf 72} (2005) 035004
  [Erratum-ibid.\ D {\bf 72} (2005) 099902];
  H.~E.~Haber and D.~O'Neil, Phys.\ Rev.\ D {\bf 74} (2006) 015018
  [Erratum-ibid.\ D {\bf 74} (2006) 059905].
\bibitem{djouadi2}A.~Djouadi, hep-ph/0503173.
\bibitem{haber2}J.~F.~Gunion and H.~E.~Haber,  Phys.\ Rev.\ D {\bf 72} (2005) 095002.
\bibitem{CP}G.C.~Branco, L.~Lavoura and S.P.~Silva, ``CP-violation'',
Oxford University Press, Oxford, England (1999).
\bibitem{ivanov}I.~P.~Ivanov,  Phys.\ Lett.\ B {\bf 632} (2006) 360.
\bibitem{nishi}C.~C.~Nishi, Phys.\ Rev.\ D {\bf 74} (2006) 036003.
\bibitem{maniatis}M.~Maniatis, A.~von Manteuffel, O.~Nachtmann and F.~Nagel,
hep-ph/0605184.
\bibitem{barroso2006}A.~Barroso, P.~M.~Ferreira, R.~Santos and J.~P.~Silva, Phys.\ Rev.\ D {\bf 74} (2006) 085016.
\bibitem{michel}L.~Michel and L.~A.~Radicati, Annals Phys.\ {\bf 66} (1971) 758.
\bibitem{Kim}J.~Kim,  Nucl.\ Phys.\ B {\bf 196} (1982) 285.
\bibitem{sartori}M.~Abud and G.~Sartori,  Phys.\ Lett.\ B {\bf 104} (1981) 147;
Annals Phys.\  {\bf 150} (1983) 307.
\bibitem{sartori2hdm}
G.~Sartori and G.~Valente, hep-ph/0304026.
\bibitem{ginzuniverse}I.~F.~Ginzburg,
  Acta Phys.\ Polon.\ B {\bf 37} (2006) 1161.
\bibitem{Branco}G.~C.~Branco and M.~N.~Rebelo,
  Phys.\ Lett.\ B {\bf 160} (1985) 117.
\bibitem{LLY}C.~l.~Lin, C.~e.~Lee and Y.~w.~Yang,
  Phys.\ Rev.\ D {\bf 50} (1994) 558.
\bibitem{GIunitarity}I.~F.~Ginzburg and I.~P.~Ivanov,
  Phys.\ Rev.\ D {\bf 72} (2005) 115010.

\end{thebibliography}
\end{document}